\documentclass[aps,reprint,pra,superscriptaddress,floats,floatfix,nobibnotes,longbibliography]{revtex4-1}

\usepackage[utf8]{inputenc}
\usepackage[T1]{fontenc}
\usepackage{amsmath}
\usepackage{graphicx,epstopdf}
\usepackage{blindtext}
\usepackage[table,xcdraw]{xcolor}
\usepackage{lipsum}
\usepackage{amsfonts}
\usepackage{bbm}
\usepackage{amssymb}
\usepackage{enumerate}
\usepackage{color}
\usepackage{latexsym}
\usepackage{amstext}
\usepackage{times}
\usepackage{threeparttable}
\usepackage{graphicx}
\usepackage[colorlinks=true,linkcolor=blue,urlcolor=blue,citecolor=blue]{hyperref}

\newcommand{\ket}[1]{| #1 \rangle}
\newcommand{\bra}[1]{\langle #1 |}
\newcommand{\ketbra}[2]{| #1 \rangle \langle #2 |}

\newcommand{\orcid}[1]{\href{https://orcid.org/#1}{\includegraphics[width=8pt]{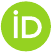}}}

\newcommand{\UFSCar}{Departamento de Computação, Universidade Federal de São Carlos, Rodovia Washington Luís, km 235 - SP-310, 13565-905 São Carlos, SP, Brazil}
\newcommand{\Unesp}{Universidade Estadual Paulista (UNESP), Instituto de Ciências e Engenharia, 18409-010 Itapeva, São Paulo, Brazil}

\begin{document}

\title{Preparation of Large Fock States in Resonators with High Probability}

\author{Lucas R. S. Santos~\orcid{0000-0001-9700-884X}}
\email[]{lucasrss@estudante.ufscar.br}
\affiliation{\UFSCar}

\author{Ciro M.~Diniz~\orcid{0000-0002-7602-0468}}
\affiliation{\UFSCar}

\author{Daniel Z. Rossatto~\orcid{0000-0001-9432-1603}}
\affiliation{\Unesp}

\author{Celso J. Villas-Boas~\orcid{0000-0001-5622-786X}}
\affiliation{\UFSCar}

\date{\today}

\begin{abstract}
    Large Fock states are important resources for bosonic quantum information and quantum-enhanced metrology, but preparing them with high probability at large excitation numbers remains challenging, as deterministic methods become increasingly control-intensive, while measurement-based approaches typically suffer from low heralding probabilities. Here we propose a protocol that combines quantum nondemolition photon-number encoding with quantum amplitude amplification to enable high-probability heralded generation of large Fock states. Starting from a cavity mode prepared in a coherent state, Quantum Phase Estimation encodes photon-number information into a multi-qubit register, while Quantum Amplitude Amplification boosts the probability of a desired target outcome before measurement. The scheme has an immediate implementation in dispersive circuit-QED, but can be analogously adapted to other bosonic platforms with QND photon-number readout, such as cavity-QED. With a register of up to eight qubits, near-deterministic preparation of Fock states with hundreds of excitations is possible. We also show that the protocol can serve as the first stage of an extension toward generating a two-mode NOON state via a conditional beam-splitter operation.
\end{abstract}

\maketitle

\section{Introduction}

Nonclassical number states of a bosonic mode, i.e., Fock states
$\ket{N}$~\cite{Hofheinz2008} are fundamental resources across quantum science and
technology. At the most basic level, these states provide perfectly
quantized field energy, serving as canonical non-Gaussian benchmarks
for calibrating the control of harmonic oscillators~\cite{Hofheinz2008,
Hofheinz2009}. They are also important resources for quantum-enhanced metrology, where number-state interferometry enables phase sensitivities approaching the Heisenberg limit~\cite{HollandBurnett1993,
Huver2008, Giovannetti2011}. A particularly relevant two-mode extension is the path-entangled NOON state, which combines number-state discreteness with mode entanglement and underlies proposals for quantum-enhanced interferometry, imaging, and
lithography~\cite{Boto2000QuantumLithography, Dowling2008HighNOON, Heras2024}.

In bosonic quantum information processing, the Fock basis provides a natural scaffold for hardware-efficient error-correcting encodings. These include binomial codes, which use superpositions of Fock states
to protect against photon loss~\cite{PhysRevX.6.031006}, and continuous-variable codes such as the Gottesman-Kitaev-Preskill (GKP) grid code~\cite{GottesmanKitaevPreskill2001}, whose
fault-tolerant operation has been demonstrated in superconducting cavities~\cite{Ofek2016Nature, CampagneIbarcq2020}. The reliable
preparation of large Fock states is also relevant for quantum simulation protocols~\cite{Wang2017} and for investigating the quantum-to-classical crossover in phase space~\cite{Heeres2015, Eickbusch2022}. An early complementary preparation route exploited interference among phase-rotated coherent states, where circular superpositions with support only on selected modular photon-number sectors can approach an individual Fock state in suitable parameter regimes~\cite{Ragi2000, Wagner2000}. While this construction provides a useful historical connection between phase-space interference and photon-number selection, it does not by itself offer a high-probability protocol for preparing a prescribed large-$N$ Fock state.

Given their wide applicability, different strategies have been developed for the preparation of large-$N$ Fock states, each with complementary requirements and performance metrics. In trapped ions, resolved sideband techniques enable deterministic preparation of motional number states~\cite{PhysRevLett.76.1796,Wagner2000,Travaglione2001}, and sequential sideband control has experimentally prepared states up to $\ket{100}$, while also demonstrating a metrological advantage~\cite{McCormick2019}. In cavity-QED, photon-number states have been generated, probed, and monitored using atomic probes and sequential quantum nondemolition (QND) measurements~\cite{Varcoe2000,PhysRevLett.88.143601,Guerlin2007,Delglise2008,Sayrin2011,SanchezMunoz18}. In waveguide QED and nanophotonic platforms, collective light-matter interactions have been explored as routes to deterministic or efficient generation of propagating multiphoton states~\cite{Tudela2015,Tudela2017}. In circuit-QED, superconducting resonators coupled to ancillary qubits provide a versatile platform for deterministic synthesis via resonant ascent and optimal control~\cite{Liu2004,Hofheinz2008,Hofheinz2009,Wang2017,Eickbusch2022,damas2025}, as well as selective number-dependent arbitrary phase (SNAP)
operations~\cite{Heeres2015, Chu2018}.

\begin{figure*}
    \centering
    \includegraphics[width = 1.0\textwidth, clip]{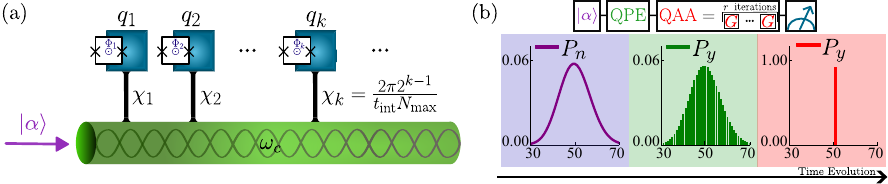}
    \caption{Circuit-QED implementation and protocol for high-probability preparation of large Fock states using quantum phase estimation (QPE) and quantum amplitude amplification (QAA). (a) Physical platform: a single bosonic mode (green) at frequency $\omega_c$ of a coplanar waveguide resonator is dispersively coupled to a register of $M$ frequency-tunable superconducting qubits, implemented as SQUID-based transmons (blue). Each qubit frequency $\omega_{q_k}(\Phi_k)$ is controlled by a local flux bias $\Phi_k$, allowing the detuning $\Delta_k(\Phi_k)=\omega_{q_k}(\Phi_k)-\omega_c$ to be calibrated. The fixed coupling $g_k$ sets the dispersive shift $\chi_k(\Phi_k)\simeq |g_k|^2/\Delta_k(\Phi_k)$ to be employed to implement the controlled powers required by QPE. A coherent pump prepares a stationary intracavity field $|\alpha\rangle$ (purple), and the dispersive interaction imprints photon-number dependent phases onto the qubit register. The controlled powers of $\hat U_{\rm QPE}$ are determined by the fixed interaction time $t_{\rm int}$ such that $\chi_k t_{\rm int}=2\pi\,2^{k-1}/N_\text{max}$, with $N_\text{max}=2^M$. (b) Protocol flow: starting from the Poissonian photon-number distribution $P_n$ of the coherent state (purple), the QPE block maps the Fock components $|n\rangle$ onto phase register outcomes $y\in\{0,\ldots,N_\text{max}-1\}$, yielding a distribution $P_y$ (green). QAA then applies the Grover operator $\hat G$ repeatedly (up to $r$ iterations) to amplify a marked outcome $y_0$, concentrating the probability in the target subspace (red). The final projective measurement of the register heralds the preparation of the target Fock state $|N\rangle$ with high probability. Shown example: $|\alpha|^2=50$, $M=7$, and targeting $|50\rangle$ ($y_0=50$).}
\label{fig:principal}
\end{figure*}

Recent work has moved beyond sequential ladder ascent by reshaping an initial coherent-state distribution through engineered interference. Coherent-control approaches combine resonant Jaynes-Cummings evolution with displacements~\cite{PhysRevLett.125.093603, Xiong2026}, repeated Kerr evolution with coherent drives~\cite{Meher2026}, Kerr-engineered Fock-space lenses~\cite{li2026, Xu2026}, or nonlinear bound states in the continuum~\cite{Rivera2023}. Resonant subspace engineering confines the oscillator dynamics to the subspace spanned by the initial coherent state and the target~\cite{Jin2026}. These methods can avoid repeated heralding and typically require only a single ancilla or no one. However, they act directly in the oscillator Hilbert space and rely on target-dependent analog parameters, optimized pulse sequences, or engineering nonlinear interactions. By contrast, our approach transfers photon-number information nondestructively to a multi-qubit register, allowing a prescribed target to be identified and marked without directly disturbing the cavity field. This QND character is closely related to cavity-QED experiments in which repeated dispersive interrogations revealed the progressive field-state collapse and photon-number quantum jumps while preserving the stored photons from direct detection~\cite{Guerlin2007}.

Measurement-based strategies based on sequential ancilla measurements and photon-number filtering have enabled the preparation of large Fock states, while adaptive generalized-parity
schemes can access larger excitation numbers with outcomes conditioned on the accumulated measurement record~\cite{PhysRevA.110.042421,
Teja2023, Deng2024, ZhangJing2026}. Related fixed-point sequences of SNAP gates and displacements amplify a target directly in the bosonic
mode~\cite{Austin2026}. By contrast, our protocol nondestructively encodes the photon number in a multi-qubit register, where a preassigned target is marked with a range and resolution set by the
register size and coherently amplified before a single final measurement.

Against this background, we address the bottleneck of reconciling high-fidelity state preparation with a high success probability at large $N$ by combining dispersive photon-number sensing with quantum amplitude amplification (QAA). In the dispersive regime, a multi-qubit register acquires photon-number-dependent phase shifts while preserving the field energy, enabling the QND encoding of the photon number into binary register outcomes via quantum phase estimation (QPE)~\cite{Schuster2007, BlaisRMP2021, kitaev_1995, nielsen2010quantum}. 
In a QPE-based, measurement-heralded approach, the success probability is ultimately limited by how much projection the initial field has into the desired photon-number component (or a small target window), so that preparing a specific large-$N$ state can require many repetitions even in the absence of control errors.
To overcome this measurement-induced probabilistic limitation, we apply the Grover-Brassard amplification primitive~\cite{Grover1996, Brassard2002} to the phase register prior to measurement. Crucially, QAA is implemented as register reflections, with the diffusion step realized by reapplying the QPE interaction (and its inverse) plus a vacuum-phase operation on the field, thereby boosting the heralding probability while leaving the conditional cavity state essentially unchanged.
In this way, the protocol replaces many incoherent experimental trials by a controlled increase in coherent circuit depth, requiring only a polynomial number of Grover iterations (scaling as $r\sim N^{1/4}$ for coherent inputs). We corroborate this scaling by numerically simulating the full open-system dynamics, including photon loss~\cite{breuer2002theory}, and show that a small number of amplification steps can substantially boost the heralding probability, enabling a scalable, near-deterministic heralded route to large-$N$ Fock-state preparation in dispersive circuit-QED. Additionally, the protocol can provide a natural starting point for
generating two-mode NOON states: after heralding $\ket{N}$ in one resonator, an auxiliary qubit can coherently control a beam-splitter exchange with a second resonator initially in the vacuum, followed by a rotated-basis qubit measurement.

The remainder of this article is organized as follows. In Sec.~\ref{sec:physicalmodel}, we introduce the circuit-QED platform and the multi-qubit Jaynes-Cummings model underlying the protocol. Section~\ref{sec:protocol} presents the QPE and QAA stages, including the dispersive phase encoding, the amplification strategy, and the heralded preparation of the target Fock state. In Sec.~\ref{sec:methods}, we describe the numerical methodology based on quantum-jump simulations of the dissipative dynamics. Section~\ref{sec:results} reports the corresponding open system performance and discusses how the prepared Fock state can be used as a resource for an extension toward two-mode NOON-state generation. In Sec.~\ref{sec:scaling}, we analyze the associated resource requirements, comparing shot overhead, circuit depth, and total execution time. Finally, Sec.~\ref{sec:conclusion} summarizes our conclusions and discusses possible extensions of the protocol. Details of the gate-level construction, the dispersive calibration, and the NOON-state extension are provided in the Appendices.

\section{Physical Model}\label{sec:physicalmodel}

Figure~\ref{fig:principal}(a) illustrates the circuit-QED platform considered throughout this work: an $M$-qubit superconducting register, implemented as frequency-tunable SQUID-based transmons and treated as effective two-level systems, dispersively coupled to a single bosonic mode of a coplanar waveguide resonator at frequency $\omega_c$. A coherent pump is used only for state preparation, generating a stationary intracavity coherent state $\ket{\alpha}$. After initialization the pump is turned off, and the subsequent coherent dynamics are governed by the multi-qubit Jaynes--Cummings Hamiltonian~\cite{JaynesCummings1963} ($\hbar=1$),
\begin{align}
\hat H&=\omega_c\,\hat a^\dagger\hat a+\sum_{k=1}^{M}\left[\frac{\omega_{q_k}}{2}\,\hat\sigma_z^{(k)} + \left(g_k \hat a\hat\sigma_+^{(k)} + \text{h.c.}\right) \right],
\label{eq:HJC_multi_main}
\end{align}
where $\hat a$ ($\hat a^\dagger$) annihilates (creates) a photon in the Fock basis of the resonator mode, which couples via $g_k$ to the $k$-th qubit (frequency  $\omega_{q_k}$), whose raising operator is $\hat\sigma_+^{(k)} = (\hat\sigma_-^{(k)})^\dagger$.

\section{Protocol}\label{sec:protocol}

Figure~\ref{fig:principal}(b) summarizes the algorithmic workflow studied here. Starting from the intracavity coherent state $\ket{\alpha}$ prepared in Sec.~\ref{sec:physicalmodel}, the QPE block maps photon-number information from the initial distribution $P_n$ onto a phase-register distribution $P_y$. The QAA block then sharpens $P_y$ by applying repeated Grover iterations prior to a final projective measurement of the register, which heralds the preparation of the target Fock state in the resonator. A detailed gate-level decomposition of the logical operations used in QPE and QAA is given in Appendix~\ref{app:gates}.

The protocol proceeds in three modules:
(i) \emph{QPE block:} after Hadamard gates preparing a uniform superposition in the register, dispersive interactions implement the controlled-phase powers required by QPE. An inverse quantum Fourier transform ($\mathrm{QFT}^\dagger$) converts the accumulated phases into a computational-basis outcome $y$ correlated with the photon number;
(ii) \emph{QAA block:} prior to measurement, Grover iterations amplify the probability of obtaining a marked outcome within the selected acceptance range, thereby increasing the heralding probability of the target Fock component without changing the measurement basis;
(iii) \emph{Measurement:} a projective measurement of the register heralds the conditional resonator state corresponding to the selected outcome, enabling preparation of the target $\ket{N}$.

\subsection{QPE block: photon-number to phase mapping}

The field is initialized in a coherent state $\ket{\alpha}$ and the qubits are prepared in the ground state $\ket{0}^{\otimes M}$. The initial state of the system is
\begin{equation}
\ket{\Psi_{\rm i}}=\ket{0}^{\otimes M}\otimes\ket{\alpha}.
\label{eq:initial_state}
\end{equation}
After applying Hadamard gates to all $M$ qubits, the register is prepared in the uniform superposition
$\left(1/\sqrt{2^M}\right)\sum_{x=0}^{2^{M}-1}\ket{x}$, which is the standard starting point of QPE: it creates an equal-weight coherent superposition of all computational basis states, allowing the subsequent evolution to encode phase information across the register amplitudes. Here $x$ labels the $M$-bit computational basis state of the register, i.e., $\ket{x}\equiv\ket{b_{M-1}\cdots b_0}$ with $x=\sum_{j=0}^{M-1} b_j 2^j$ and $b_j\in\{0,1\}$.

The QPE block then maps photon-number information onto the qubit register through the photon-number-dependent phases generated by the dispersive interaction. It is therefore convenient to introduce the normalized unitary phase operator
\begin{equation}
\hat U_{\rm QPE}=\exp\!\left(-i\,\frac{2\pi}{N_\text{max}}\,\hat n\right),
\label{eq:U_qpe}
\end{equation}
where $\hat n\equiv \hat a^\dagger\hat a$ is the cavity photon-number operator. The respective eigenstates are the Fock states $\ket{n}$, with eigenvalues $\exp(-i2\pi n/N_\text{max})$. Thus, $N_\text{max}=2^M$ sets the phase resolution of the $M$-qubit register, namely, the number of distinguishable phase values that can be encoded and later resolved by the inverse quantum Fourier transform. In our circuit-QED implementation, the controlled powers associated with this unitary arise from dispersive phase accumulation, as discussed below, and their calibration is derived in Appendix~\ref{app:uqpe_from_hint}.

To formally connect Eq.~\eqref{eq:U_qpe} to the physical model, we consider the dispersive regime on Eq.~\eqref{eq:HJC_multi_main},
$|\Delta_k|\equiv|\omega_{q_k}-\omega_c|\gg g_k\sqrt{\bar n+1}$, where $\bar n=\langle \hat n\rangle$ is the intracavity photon number. As detailed in Appendix~\ref{app:uqpe_from_hint}, in this limit we perform the standard Schrieffer–Wolff (small-rotation) transformation $\hat U=\exp[(g_k/\Delta_k)(\hat a\hat\sigma_+-\hat a^\dagger\hat\sigma_-)]$ and truncate the Baker–Campbell–Hausdorff expansion to second order in $g_k/\Delta_k$~\cite{JaynesCummings1963,BlaisRMP2021}. Under these conditions, working in the interaction picture, we obtain the dispersive interaction 
\begin{equation}
\hat{H}_{\rm int}=\sum_{k=1}^{M}\chi_k\,\hat n \hat{\sigma}_z^{(k)},
\label{eq:Hint}
\end{equation}
with dispersive shifts $\chi_k\simeq |g_k|^2/\Delta_k$. 

Equation~\eqref{eq:Hint} shows that, during an interaction window, the register acquires photon-number-dependent phases without energy exchange. In particular, for qubit $k$ of a two-qubit register, prepared in the superposition $\ket{+}_k=(\ket{0}_k+\ket{1}_k)/\sqrt{2}$ and the field in the Fock component $\ket{n}$, the dispersive evolution for an arbitrary time $t_{\rm int}$ yields
$\ket{+}_k\ket{n}\mapsto(\ket{0}_k+e^{-i\chi_k n t_{\rm int}}\ket{1}_k)\ket{n}/\sqrt{2}$.
To implement the $k$-th controlled power in QPE, we match this physical phase to the algorithmic action of Eq.~\eqref{eq:U_qpe}, $\hat U_{\rm QPE}^{2^{k-1}}\ket{n}=\exp[-i(2\pi\,2^{k-1}/N_\text{max})\,n]\ket{n}$,
which fixes the per-photon calibration as
\begin{equation}
\chi_k\,t_{\rm int}=\frac{2\pi\,2^{k-1}}{N_\text{max}}.
\label{eq:calib_cond}
\end{equation}
For the full derivation, see Appendix~\ref{app:uqpe_from_hint}. Since $\hat U_{\text{QPE}}$ is generated by $\hat n$, raising it to the power $2^{k-1}$ only rescales the phase, so each controlled power is implemented through phase matching rather than by $2^{k-1}$ sequential applications.

In the circuit-QED implementation considered here, we adopt a parallel scheme in which all register qubits interact with the resonator during a common time window $t_{\rm int}$, while the binary weighting is encoded in qubit-dependent dispersive shifts. We assume that the couplings $g_k$ are fixed by device design, whereas the qubit frequencies are tunable. The required shifts are obtained by flux-biasing each qubit to a detuning $\Delta_k(\Phi_k)=\omega_{q_k}(\Phi_k)-\omega_c$~\cite{Krantz2019,Kjaergaard2020}, which sets, within the dispersive approximation, $\chi_k(\Phi_k)\simeq |g_k|^2/\Delta_k(\Phi_k)$. Higher-order corrections and the multilevel structure of transmons can be absorbed into the experimentally calibrated value of $\chi_k$. Thus, in the baseline realization analyzed below, the controlled powers required by QPE are implemented by tuning qubit detunings at a fixed interaction time.

The calibration above relies on the dispersive approximation, remaining accurate over the photon-number manifold populated by the coherent input. This description is controlled by the small parameter
$\left(g_k/\Delta_k\right)\sqrt{\bar n+1}$~\cite{Schuster2007}: when the intracavity photon number $\bar{n}$ approaches the critical photon number $n_{\rm crit}\simeq \Delta_k^{2}/(4g_k^{2})$, higher-order corrections and the eventual breakdown of the dispersive expansion become non-negligible~\cite{BlaisRMP2021}. Since the input is a coherent state with $\bar n=N$ and Poissonian width $\sqrt{N}$, we require $n_{\rm crit}\gtrsim \bar n_\text{max}\equiv N+s\sqrt{N}$ (with $s\sim4$--$5$) at the largest targets. Representative circuit-QED parameters $g_k/2\pi\sim 10$--$100~\mathrm{MHz}$ and $\Delta_k/2\pi\sim 1$--$3~\mathrm{GHz}$~\cite{Wallraff2004} yield $n_{\rm crit}\gtrsim 10^3$, while dispersive shifts
$|\chi_k|/2\pi\sim 0.1$--$1~\mathrm{MHz}$ correspond to interaction times
$t_\pi\equiv \pi/|\chi_k|\approx 0.5$--$5~\mu\mathrm{s}$ to accumulate a conditional $\pi$ phase, compatible with current decoherence times in state-of-the-art circuit-QED devices~\cite{Krantz2019, Kjaergaard2020}. In the numerical simulations discussed below, we set $\kappa t_{\rm int}=10^{-3}$, so that the dispersive encoding occurs well within the cavity lifetime.

The role of QPE is to convert the photon-number dependent phases imprinted across the register superposition into a binary readout $y$ by applying controlled powers of $\hat U_{\rm QPE}$ followed by the inverse quantum Fourier transform $\mathrm{QFT}^\dagger$~\cite{cleve1998quantum,Torosov2009,nielsen2010quantum}. Concretely, after the controlled-$\hat U_{\rm QPE}$ stage, the register amplitudes in the computational basis $\{\ket{x}\}$ acquire a phase gradient of the form $\exp(-i2\pi n x/N_\text{max})$ for each Fock component $\ket{n}$. The operation $\mathrm{QFT}^\dagger$ performs the corresponding discrete Fourier decoding, i.e., it changes the register basis from $\ket{x}$ to $\ket{y}$ and converts this phase information into a computational-basis amplitude concentrated near an integer outcome $y$ that estimates the encoded phase. Measuring the qubits after $\mathrm{QFT}^\dagger$ therefore yields an integer $y\in\{0,\ldots,N_{\max}-1\}$, which resolves the photon number modulo $N_{\max}$. Consequently, $\ket{N}$ and $\ket{N+\ell N_{\max}}$ (with integer $\ell$) are indistinguishable under this encoding. To minimize aliasing errors from the Poissonian tail of the input $\ket{\alpha}$, we choose $N_\text{max}$ to cover the relevant distribution support (and in our simulations we truncate the cavity Hilbert space at this cutoff). In this regime, the measurement produces a phase distribution $P_y$ strongly correlated with $n$, so that obtaining a given outcome $y$ heralds a conditional collapse of the field toward the corresponding photon-number component.

To target a desired Fock state $\ket{N}$ (within the representable range $N\le N_\text{max}$), we identify the nominal register outcome $y_0$ associated with $N$ on the QPE phase grid. When the relevant photon-number support of the input state lies within the chosen cutoff $N_\text{max}$, the QPE readout is sharply correlated with photon number, so that the target component $\ket{N}$ is associated with the outcome $y_0=N$. To increase robustness against finite broadening of the measured distribution, we may accept a small subspace of half-width $w$ around $y_0$, i.e., ${\rm good}=\{y_0-w,\ldots,y_0+w\}$ (understood modulo $N_\text{max}$). Enlarging this window increases the heralding probability at the cost of accepting nearby photon-number components. In general, we write the projector onto the good subspace as
\begin{equation}
\hat \Pi_{\rm good}=\sum_{y\in{\rm good}} \ketbra{y}{y} \otimes \hat I_f,
\label{eq:proj_good}
\end{equation}
where $\hat I_f$ represents the identity operator on the field Hilbert space. Denoting $\ket{\Psi_0}$ as the joint state after the QPE block, explicitly (as described in Appendix~\ref{app:gates}), the unmeasured QPE output can be written as
\begin{equation}
    \ket{\Psi_0}=(\mathrm{QFT}^\dagger\otimes \hat I_f)\,
\frac{1}{\sqrt{2^M}}\sum_{x=0}^{2^M-1}\ket{x}\otimes \hat U_{\rm QPE}^{x}\ket{\alpha},
\end{equation}
so that, before applying $\mathrm{QFT}^\dagger$, each Fock component $\ket{n}$ imprints a phase gradient $\exp(-i2\pi n x/N_\text{max})$ across the register amplitudes $\ket{x}$. The inverse quantum Fourier transform then converts this phase pattern into amplitudes in the measurement basis $\ket{y}$, yielding a distribution $P_y$ peaked around the value associated with that photon-number component. The initial success probability (i.e., the probability that a measurement yields $y\in{\rm good}$) is then
\begin{equation}
a_0=\bra{\Psi_0}\hat \Pi_{\rm good}\ket{\Psi_0}=P(y\in{\rm good}).
\end{equation}

\subsection{QAA block: amplifying the marked phase outcome}

When targeting large-$N$ states from a coherent resource $|\alpha|^2=N$, the baseline success probability after the QPE block is limited by Poissonian number statistics. Concretely, $a_0$ is the probability that the register measurement yields an accepted outcome using QPE alone (i.e., before any amplification). For a single marked outcome $y_0$ corresponding to $N$, this reduces to the initial weight of $\ket{N}$ in the coherent state. Using Stirling's approximation~\cite{AbramowitzStegun},
\begin{equation}
a_0 \simeq P_n(N)=e^{-N}\frac{N^N}{N!}\approx \frac{1}{\sqrt{2\pi N}}.
\label{eq:a0_QPE}
\end{equation}
Thus, without QAA the expected number of incoherent repetitions required to herald success scales as $\mathcal{O}(1/a_0)$ (i.e., $\mathcal{O}(N^{1/2})$ for coherent inputs).

Instead, we apply QAA to amplify the probability of obtaining $y\in{\rm good}$ prior to measurement (the gate-level construction is detailed in Appendix~\ref{app:gates}). The post-QPE joint state can be decomposed as
\begin{subequations}\label{eq:good_bad_decomp}
\begin{align}
&\ket{\Psi_0} = \sqrt{a_0}\,\ket{\Psi_{\rm good}}+\sqrt{1-a_0}\,\ket{\Psi_{\rm bad}}, \label{eq:good_bad_decomp_a}\\
&\ket{\Psi_{\rm good}} = \frac{1}{\sqrt{a_0}}\,\hat \Pi_{\rm good}\ket{\Psi_0}, \label{eq:good_bad_decomp_b}\\
&\ket{\Psi_{\rm bad}} = \frac{1}{\sqrt{1-a_0}}\left(\hat I-\hat \Pi_{\rm good}\right)\ket{\Psi_0}, \label{eq:good_bad_decomp_c}
\end{align}
\end{subequations}
where $\ket{\Psi_{\rm good}}$ and $\ket{\Psi_{\rm bad}}$ are normalized components supported on the accepted and rejected subspaces, respectively. The operator $\hat I$ denotes the identity operator on the full register-field Hilbert space.

The QAA step consists of repeated applications of the Grover iterate $\hat G=\hat S_{\Psi}\hat S_{\chi}$~\cite{Brassard2002}, where $\hat S_{\chi}$ marks the accepted register outcomes $y\in{\rm good}$ and $\hat S_{\Psi}$ reflects about the joint state prepared by the QPE block. In Appendix~\ref{app:gates}, a circuit-QED motivated gate decomposition of these reflections is provided, highlighting that each Grover iterate can be implemented using one application of the QPE block and one of its inverse, together with a vacuum-selective cavity phase~\cite{Schuster2007, Heeres2015, Krastanov2015} and standard multiqubit controlled-phase logic~\cite{Barenco1995, Glaser2023, Chen2024}.

Each application of $\hat G$ performs a rotation in the two-dimensional subspace spanned by $\{\ket{\Psi_{\rm good}},\ket{\Psi_{\rm bad}}\}$. Defining $\sin^2\theta=a_0$, after $r_y$ Grover iterations the state becomes $\ket{\Psi_{r_y}}=\sin\!\big[(2r_y+1)\theta\big]\ket{\Psi_{\rm good}}+\cos\!\big[(2r_y+1)\theta\big]\ket{\Psi_{\rm bad}}$~\cite{Brassard2002}, and the corresponding success probability is
\begin{equation}
    P_{\rm suc}(r_y)=\sin^2\!\big[(2r_y+1)\theta\big].
    \label{eq:P_suc_f}
\end{equation}
Maximal success ($P_{\rm suc}\approx1$) occurs when $(2r_y+1)\theta\approx\pi/2$. This yields the optimal iteration count
\begin{equation} 
r_y=\frac{\pi}{4\arcsin\big(\sqrt{a_0}\big)}-\frac{1}{2}\approx \frac{\pi}{4\sqrt{a_0}}-\frac{1}{2}, 
\label{eq:ry} 
\end{equation} 
which, using Eq.~\eqref{eq:a0_QPE}, implies the favorable scaling $r_y\propto N^{1/4}$. The final step is a projective measurement of the qubit register. By choosing $r_y$ near the optimum, QAA can raise the heralding probability close to unity, so that the protocol becomes effectively deterministic. Upon obtaining a successful outcome $y\in{\rm good}$, the conditional cavity state is prepared. For ${\rm good}=\{y_0\}$ it approaches the target Fock component provided the QPE grid is sufficient, i.e., $N_{\max}$ is large enough that aliasing from photon-number components outside the representable range is negligible. To assess how this deterministic preparation strategy performs in the presence of cavity loss, we now turn to stochastic simulations of the full protocol.

\section{Methods}
\label{sec:methods}

To study the protocol efficiency in the presence of dissipation, we use the Monte Carlo method~\cite{Molmer93, Johansson2012},
which provides a stochastic pure-state trajectory of the Lindblad dynamics associated with cavity photon loss~\cite{breuer2002theory}. During the dispersive phase-accumulation intervals, the corresponding master equation is
\begin{equation}
    \frac{d\hat\rho}{dt}
    =
    -i[\hat H_{\rm int},\hat\rho]
    +
    \kappa
    \left(
        \hat a\hat\rho\hat a^\dagger
        -
        \frac{1}{2}
        \left\{
            \hat a^\dagger\hat a,\hat\rho
        \right\}
    \right),
    \label{eq:master_methods}
\end{equation}
with the jump operator $\hat L=\sqrt{\kappa}\hat a$. Rather than directly integrating Eq.~\eqref{eq:master_methods}, we simulate its quantum-jump unraveling. Between successive jumps, each trajectory is propagated according to the effective non-Hermitian Schrödinger
equation
\begin{equation}
    i\frac{d}{dt}\ket{\tilde{\psi}(t)}
    =
    \left(
        \hat H_{\rm int}
        -
        \frac{i}{2}\hat L^\dagger\hat L
    \right)
    \ket{\tilde{\psi}(t)},
    \label{eq:H_NH_methods}
\end{equation}
where $\ket{\tilde{\psi}(t)}$ denotes the unnormalized no-jump state. For a normalized trajectory state $\ket{\psi(t)}$, a photon-loss event occurs during $dt$ with probability
$dp_{\rm jump}=dt\,\bra{\psi}\hat L^\dagger\hat L\ket{\psi}$, after which
$\ket{\psi}\rightarrow \hat L\ket{\psi}/
\sqrt{\bra{\psi}\hat L^\dagger\hat L\ket{\psi}}$.
Averaging over stochastic trajectories recovers the Lindblad dynamics
of Eq.~\eqref{eq:master_methods}.

We quantify the performance of the protocol by simulating the full sequence described in Sec.~\ref{sec:protocol} for representative target photon numbers. The next section will focus on the narrow acceptance case ${\rm good}=\{y_0\}$ (i.e., window width $w=0$), so that QAA amplifies a single marked register outcome associated with the target Fock component $\ket{N}$. To maximize the initial QPE overlap with the target, we choose the coherent input amplitude such that $|\alpha|^2=N$. We also assume the commensurate setting $N_\text{max}=2^M$ and truncate the cavity Hilbert space at the same cutoff, so that within the simulated manifold the QPE readout provides a one-to-one correspondence between photon numbers and computational basis outcomes ($y=n$ for $n\in\{0,\ldots,N_\text{max}-1\}$), while suppressing aliasing from photon-number components beyond the truncation.

\section{Results}
\label{sec:results}

\begin{figure}[htbp]
    \centering
    \includegraphics[width = 0.48\textwidth, clip]{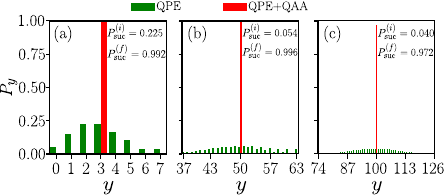}
    \caption{Register phase distribution before and after amplification for target Fock states: (a) $M=3$, $N=3$; (b) $M=6$, $N=50$; and (c) $M=7$, $N=100$. In each panel, the green bars show the QPE distribution $P_y^{(i)}$, while the red bars show the amplified distribution $P_y^{(f)}$ obtained after the QPE+QAA sequence. The resonator is initialized in a coherent state with $|\alpha|^2=N$, and the acceptance set is restricted to a single marked outcome, ${\text{good}}=\{y_0\}$, associated with the target $\ket{N}$. The values reported in each panel correspond to the target weights $P_y^{(i)}(y_0)$ and $P_y^{(f)}(y_0)$, i.e., the initial and final success probabilities for heralding the target Fock component.}
    \label{fig:qpe_qaa_distributions}
\end{figure}

Under the numerical conditions described in Sec.~\ref{sec:methods}, we first compare the register distributions obtained only using QPE and after the full sequence of QPE and QAA for representative targets $N=3$, $N=50$, and $N=100$, as shown in Fig.~\ref{fig:qpe_qaa_distributions}. The success probability presented is the probability weight at the target outcome, i.e., $P_{\rm suc}=P_y(y_0)$. The unamplified distributions $P_y^{(i)}$ reflect the Poissonian weight of the coherent input and therefore decrease with increasing $N$, in agreement with the scaling of Eq.~\eqref{eq:a0_QPE}. By contrast, the amplified distributions $P_y^{(f)}$ show that QAA transfers most of the probability weight to the marked outcome $y_0$, yielding a pronounced concentration of the register distribution around the target value, as quantified by the target weights $P_y^{(i)}(y_0)$ and $P_y^{(f)}(y_0)$ shown in each panel.

This enhancement is already substantial for small targets and becomes especially significant in the large-$N$ regime, where QPE alone would require many repetitions to herald the desired Fock component. For example, for $N=100$, the initial target weight is only of order $P_\text{suc}^{(i)} = 0.040$, whereas after amplification the probability at $y_0$ approaches unity. In this sense, QAA converts the algebraically decaying baseline probability of the coherent input into an effectively deterministic heralding scheme, while the amplification overhead remains limited by the favorable scaling $r_y\propto N^{1/4}$. Because the amplification acts on the register subspace, the conditional fidelity of the field state remains high, so the success probability at the marked outcome serves as the main performance metric.

\begin{figure}[htbp]
    \begin{centering}
    \includegraphics[width = 0.48\textwidth, clip]{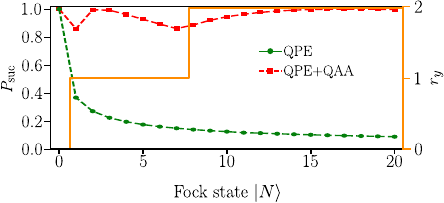}
    \par\end{centering}
    \caption{Success probability for preparing $\ket{N}$ versus target photon number for a fixed register size ($M=5$). The left axis shows the QPE and amplified (QPE+QAA) success probabilities, $P_{\text{suc}}^{(i)}=P_y^{(i)}(y_0)$ and $P_{\text{suc}}=P_y^{(f)}(y_0)$, respectively, for the narrow acceptance case ${\text{good}}=\{y_0\}$. The right axis shows the selected number of Grover iterations $r_y$ used in the amplification step.}
    \label{fig:psuc_ry}
\end{figure}

To elucidate how amplification behaves across targets for a fixed register size, Fig.~\ref{fig:psuc_ry} shows the success probability $P_{\text{suc}}$ (left axis) together with the chosen integer $r_y$ (right axis) as functions of $N$ for $M=5$. As expected from Eq.~\eqref{eq:a0_QPE}, the QPE success probability decreases with increasing $N$. After QAA, however, $P_{\text{suc}}$ remains close to unity over the full accessible range, while the optimal number of Grover steps grows only in discrete increments. The small valleys visible in the amplified success probability are a finite-integer effect: the exact maximum of Eq.~\eqref{eq:P_suc_f} generally occurs at a non-integer value of $r_y$, so in practice we choose the nearest integer (floor or ceiling) that maximizes the post-amplification probability.

\begin{figure}[htbp]
    \begin{centering}
    \includegraphics[width = 0.49\textwidth, clip]{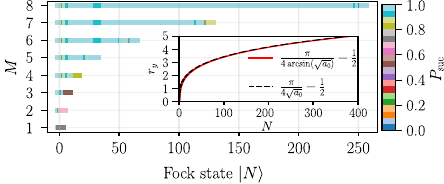}
    \par\end{centering}
    \caption{Fock-state generation map for the QPE+QAA protocol. Each marker corresponds to a target $\ket{N}$ (horizontal axis) addressed with an $M$-qubit register (vertical axis). The color scale indicates the post-amplification success probability $P_{\text{suc}}=P_y^{(f)}(y_0)$ for the narrow acceptance case ${\text{good}}=\{y_0\}$. Inset: estimated Grover-iteration number $r_y(N)$ from the exact expression (solid) and its small-$a_0$ approximation (dashed).}
    \label{fig:fock_map}
\end{figure}

Finally, Fig.~\ref{fig:fock_map} summarizes the accessible photon-number manifold as a function of the number of qubits $M$. Since QPE resolves phases on a grid of size $N_\text{max}=2^M$, increasing the register size exponentially enlarges the set of distinguishable Fock targets. The color scale shows that high post-amplification success probabilities are maintained across this enlarged manifold. The inset compares Eq.~\eqref{eq:ry} with its large-$N$ approximation obtained from $\arcsin(\sqrt{a_0})\simeq \sqrt{a_0}$ for $a_0\ll1$, confirming that the amplification overhead $r_y$ grows only as $N^{1/4}$. In practice, this means that even for photon numbers in the thousands only a modest number of Grover iterations is required. For example, with $M=10$ the target state $\ket{1000}$ is reached with only about $r_y\approx 6$ iterations.

In addition to its direct use as a single-mode nonclassical resource, the high-probability preparation of large Fock states also suggests a route toward two-mode entangled photon-number states. A standard example is the NOON state,
\begin{equation}
    \ket{\mathrm{NOON}}_{N,\phi}
    =
    \frac{1}{\sqrt{2}}
    \left(
    \ket{N}_a\ket{0}_b
    +
    e^{i\phi}\ket{0}_a\ket{N}_b
    \right),
\end{equation}
which is a useful resource for quantum-enhanced interferometry~\cite{Boto2000QuantumLithography,Dowling2008HighNOON,Huver2008,Giovannetti2011}. One could imagine extending the presented protocol directly to two coherent modes and amplifying the joint subspace. However, as shown in Appendix~\ref{app:noon_extension}, this direct projection is exponentially unfavorable, leading to an amplitude amplification overhead $r^{\rm NOON}_y\propto 2^{N/2}N^{1/4}$.

A more promising extension is therefore to use the present protocol only to prepare the state $\ket{N}$ in a first cavity mode, add a second mode initialized in the vacuum, and then perform a coherent conditional mode-exchange operation. An auxiliary qubit is prepared
in $\ket{+}_q=(\ket{0}_q+\ket{1}_q)/\sqrt{2}$ and controls an exchange between the two resonators. The required mode exchange is generated by the beam-splitter Hamiltonian
\begin{equation}
    \hat H_{\rm BS}(t)
    =g_{\rm BS}(t)
    \left(
        e^{i\varphi_{\rm BS}}\hat a^\dagger\hat b
        +
        e^{-i\varphi_{\rm BS}}\hat a\hat b^\dagger
    \right),
    \label{eq:H_bs_main}
\end{equation}
for which a pulse area
$\Theta=\int_0^{t_{\rm BS}}g_{\rm BS}(t)\,dt=\pi/2$ completely transfers $\ket{N,0}_{ab}$ into $\ket{0,N}_{ab}$, up to a known phase. At the effective level, the conditional operation required here is described by
\begin{equation}
    \hat H_{\rm control-BS}(t)
    =
    \ket{1}_q\bra{1}\otimes\hat H_{\rm BS}(t),
    \label{eq:H_cbs_main}
\end{equation}
so that the exchange is applied only to the $\ket{1}_q$ branch, while the $\ket{0}_q$ branch remains unchanged. The resulting state has the
form
\begin{equation}
    \frac{1}{\sqrt{2}}
    \left(
        \ket{0}_q\ket{N,0}_{ab}
        +
        e^{i\Phi}\ket{1}_q\ket{0,N}_{ab}
    \right),
\end{equation}
where $\Phi$ contains the controllable qubit phase and the deterministic phase accumulated during the swap. A final measurement of the
auxiliary qubit in the $\ket{\pm}$ basis erases the which-mode information and heralds the corresponding NOON state.

More generally, atom-mediated two-mode cavity-QED schemes provide a microscopic route to such state-dependent exchanges: driven three-level and two-level atoms can be engineered to generate effective bilinear interactions of the beam-splitter, with the effective coupling
conditioned by the atomic state~\cite{VillasBoas2005, Prado2006}. Parametrically activated beam-splitter interactions of the form in Eq.~\eqref{eq:H_bs_main}, including fast high-fidelity swaps between superconducting cavities, have also been experimentally demonstrated using Josephson nonlinear couplers~\cite{Lu2023HighFidelityBeamsplitter}. Moreover,
controlled-SWAP (Fredkin) operations between two microwave cavity modes have been realized with a transmon ancilla, providing a direct circuit-QED route to the conditional exchange required here~\cite{Gao2019}. Related transmon/qutrit-mediated protocols have also been proposed specifically for microwave NOON-state
generation~\cite{Su2014NOONCircuitQED}. The detailed algebra, phase conventions, and possible phase correction are given in Appendix~\ref{app:noon_extension}.

\section{Resource scaling}
\label{sec:scaling}

The utility of the protocol is governed by two distinct resources: the number of independent repetitions (shots) and the time budget of each shot. Since the register size grows only logarithmically with the target, $M\simeq \log_2 N$, the key question is whether the quadratic reduction in repetition count provided by QAA compensates for the increased depth of a single execution.

Without amplification, each execution of the protocol (state preparation + QPE + register readout) is an independent trial with success probability $a_0$ for obtaining $y\in{\rm good}$. Hence, the probability of zero success after $\eta$ shots is $(1-a_0)^\eta$, and requiring at least one success with desired probability $p_d$, i.e.\ $1-(1-a_0)^\eta \ge p_d$, yields
\begin{equation}
\eta\ge \frac{\ln(1-p_d)}{\ln(1-a_0)}.
\end{equation}
For coherent inputs with $a_0\sim N^{-1/2}$, this repetition overhead scales as $\mathcal{O}(1/a_0)\sim \mathcal{O}(N^{1/2})$. As shown in Table~\ref{tab:qpe_qaa_summary}, for $N=200$ and $p_d=0.95$ this probabilistic strategy requires more than $10^2$ repetitions.

\begin{table}[b!]
\caption{\label{tab:qpe_qaa_summary} Resource summary for representative targets $\ket{N}$.
$M$ is the number of qubits and $a_0$ is the QPE success probability.
$\eta$ is the number of repetitions required without amplification to reach $p_d=0.95$.
With QAA, $r_y$ is the number of Grover iterations and $P_{\text{suc}}$ is the resulting success probability in a single run.}
\begin{ruledtabular}
\begin{tabular}{c c c c c c}
$|N\rangle$ & $M$ & $a_0$ & $\eta$ & $r_y$ & $P_{\rm suc}$ \\
\hline
$|3\rangle$ & 3 & 0.225 & 12 & 1 & 0.992 \\
$|50\rangle$ & 6 & 0.054 & 55 & 3 & 0.996 \\
$|100\rangle$ & 7 & 0.040 & 74 & 3 & 0.973  \\
$|200\rangle$ & 8 & 0.028 & 105 & 4 & 0.997 \\
\end{tabular}
\end{ruledtabular}
\end{table}

With QAA, the success probability is instead increased within a single coherent run. The optimal number of Grover iterations scales as $r_y\propto N^{1/4}$, so the number of required QPE-based attempts drops from $\mathcal{O}(1/a_0)$ to $\mathcal{O}(1/\sqrt{a_0})$. Although this increase in the number of repetitions could be hindered by circuit depth, the comparison of execution times shows the opposite. Beyond requiring fewer shots, the QAA protocol has a shorter execution time than many repetitions of the QPE protocol. 
As detailed in Appendix~\ref{app:gates}, the Grover operator can be written as $\hat G=\hat{\mathcal A}\,\hat S_0\,\hat{\mathcal A}^\dagger\,\hat S_\chi$, where $\hat{\mathcal A}$ denotes the full state-preparation block (coherent displacement, register initialization, QPE phase encoding, and $\mathrm{QFT}^\dagger$), $\hat S_\chi$ is the oracle reflection on the marked register subspace, and $\hat S_0$ is the reflection about the initialized reference state, which is compiled from a multi-controlled phase flip on the qubit register together with a vacuum-selective SNAP phase on the cavity~\cite{Schuster2007, Heeres2015, Krastanov2015}. Therefore, each Grover iteration contains one application of $\hat{\mathcal A}$ and one of $\hat{\mathcal A}^\dagger$, in addition to the two reflections. Hence, the total logical depth then scales as
\begin{equation}
D_{\rm tot}\approx (2r_y+1)\,D_{\hat{\mathcal A}}+r_y\,(D_{\hat S_\chi}+D_{\hat S_0}),
\end{equation}
where $D_{\hat{\mathcal A}}$ is the depth of the full preparation (QPE block), $D_{\hat S_\chi}$ is the depth of the register oracle, and $D_{\hat S_0}$ is the depth of the reference-state reflection. In other words, a single amplified shot contains $2r_y+1$ applications of the block $\hat{\mathcal A}$. For the largest target studied here, $N=200$, we have $r_y=4$, so one amplified execution requires $2r_y+1=9$ uses of $\hat{\mathcal A}$.

An experimental benchmark is the work of Deng et al.~\cite{Deng2024}, who generated large Fock states by sequentially re-preparing a single dispersively coupled ancilla to implement sinusoidal photon number filters (PNF). Their approach is less costly in simultaneous qubit hardware and has a circuit depth $d=\log_2\sqrt{N}$. For $\ket{100}$, the experimental sequence employed two PNF assisted by a Gaussian filter. Nevertheless, the probability of selecting a target Fock state from an initial coherent state remains limited by its Poissonian weight, $P_N\simeq(2\pi N)^{-1/2}$, giving $P_{100}\simeq0.040$. Under the $p_d=0.95$ confidence criterion adopted in
Table~\ref{tab:qpe_qaa_summary}, this ideal Poisson-limited probability corresponds to 74 independent preparation attempts. The experimentally reported value $P_{100}=0.55(2)$ instead denotes the target population within the postselected state, rather than its heralding probability, and the corresponding spectroscopy characterization accumulated approximately $10^6$ total trials for $2.1\times10^3$ postselected records. By contrast, our open-system simulations predict that a 7-qubit QPE register followed by three Grover iterations raises the single-run heralding probability to $P_{\rm suc}=0.973$. These iterations coherently reuse the QPE preparation map and its inverse, but require no intermediate measurements and only one
final register readout. The comparison therefore exposes a hardware-sampling trade-off: the PNF protocol uses one recycled ancilla but relies on sequential measurement, feedback, and postselection, whereas the protocol presented in this work employs a logarithmically growing register, $M\simeq\log_2N$, to replace many probabilistic preparations with one deeper coherent execution.

Then, to make contact of this protocol with physical feasibility, one must further translate this logical depth into coherent time. Denoting by $T_{\hat{\mathcal{A}}}$, $T_{\hat S_\chi}$, and $T_{\hat S_0}$ the corresponding operation times, the total coherent duration of one amplified shot is
\begin{equation}
T_{\rm tot}\approx (2r_y+1)\,T_{\hat{\mathcal A}}+r_y\,(T_{\hat S_\chi}+T_{\hat S_0}).
\label{eq:Ttot_scaling}
\end{equation}
Under the parallel implementation assumed in Sec.~\ref{sec:protocol}, the controlled powers are realized simultaneously by calibrating the detuning-dependent dispersive shifts $\chi_k(\Phi_k)$ at a common interaction time $t_{\rm int}$. The calibration condition of Eq.~\eqref{eq:calib_cond} shows that the largest controlled phase corresponds to the largest register weight, for which $\chi_{\max}t_{\rm int}\sim\pi$. Therefore, the QPE interaction window is set by the largest dispersive shift that can be calibrated while remaining safely in the dispersive regime. For representative strong-dispersive values $|\chi_{\max}|/2\pi\sim 1~\mathrm{MHz}$~\cite{BlaisRMP2021}, one obtains $t_{\rm int}\sim\pi/|\chi_{\max}|\sim 0.5~\mu\mathrm{s}$, well below the coherence times relevant for superconducting-resonator state manipulation~\cite{Hofheinz2009, Reagor2016, Milul2023}. Once this reference interaction time is fixed, the remaining dispersive shifts are determined directly from Eq.~\eqref{eq:calib_cond}.

In the detuning-calibrated realization considered here, the binary weighting is obtained by tuning the qubit frequencies, so that $\chi_k(\Phi_k)\simeq |g_k|^2/\Delta_k(\Phi_k)$. For identical fixed couplings $g_k$, a fully parallel implementation requires the dispersive shifts, and therefore the inverse detunings, to span a range of order $2^{M-1}$. This is a hardware constraint rather than an algorithmic one, and it can be relaxed by device-level choices of nonidentical $g_k$~\cite{BlaisRMP2021, Krantz2019, Kjaergaard2020}, by restricting to moderate register sizes, or by using more elaborate superconducting architectures such as tunable-coupler designs~\cite{Allman2010, wulschner2016, Glaser2023}. For instance, a parallel circuit-QED alternative would be to calibrate the dispersive shifts through effective qubit--resonator couplings, $g_k\rightarrow g^{\rm eff}_k(\Phi_{c,k})$, using flux-controlled tunable couplers~\cite{Allman2010,Glaser2023}. In that case, part of the calibration burden is transferred from the qubit--resonator detuning range to the tunability range of the effective coupling, since $\chi_k$ can be programmed through $|g^{\rm eff}_k(\Phi_{c,k})|^2/\Delta_k$. By contrast, a time-programmed implementation could keep a common dispersive shift $\chi$ and instead use qubit-dependent interaction windows, $t_k=2\pi 2^{k-1}/N_{\max}\chi$. This strategy is more closely aligned with QND photon counting in cavity-QED, where probe atoms acquire photon-number-dependent phases during controlled dispersive interactions with the cavity field~\cite{Varcoe2000, PhysRevLett.88.143601, Guerlin2007, Sayrin2011}.

With this implementation choice fixed, the remaining contribution to $T_{\hat{\mathcal A}}$ comes from the register logic within $\hat{\mathcal A}$, namely the Hadamard gates and the $\mathrm{QFT}^\dagger$. These operations rely on single- and two-qubit gates with typical durations of $\sim 20~\text{ns}$ and $\sim 100~\text{ns}$, respectively \cite{Krantz2019, Kjaergaard2020}. In a fully connected architecture, the critical path of a parallelized $M$-qubit $\mathrm{QFT}^\dagger$ scales as $\mathcal{O}(M)$, requiring approximately $M$ layers of two-qubit gates and $M$ layers of single-qubit gates \cite{cleve1998quantum}. For an $M=8$ register ($N=200$), this optimized logic sequence takes $8 \times 100~\text{ns} + 8 \times 20~\text{ns} \approx 1.0~\mu\text{s}$. Adding this logical delay to the dispersive interaction window, the total duration of the state-preparation block becomes $T_{\hat{\mathcal{A}}} \approx 1.5~\mu\text{s}$. The oracle reflection $\hat S_\chi$ consists of a multi-controlled phase flip on the qubits, which can be compiled efficiently in modern circuit-QED architectures to take $T_{\hat S_\chi} \approx 0.5$--$1.0~\mu\text{s}$ \cite{Krantz2019}. Finally, the reference reflection $\hat S_0$ includes a vacuum-selective SNAP operation on the cavity, typically requiring $T_{\hat S_0} \approx 1.0$--$2.0~\mu\text{s}$ \cite{Heeres2015, Eickbusch2022}.

For the largest target studied here, $N=200$ ($r_y=4$), the full amplified execution requires $9$ applications of $\hat{\mathcal{A}}$ and $4$ applications of both $\hat S_\chi$ and $\hat S_0$. Summing these contributions gives a single-run coherent duration of $T_{\text{tot}} \approx 20$--$25~\mu\text{s}$. While this operation must be completed within the system coherence limit, it remains safely bounded by the millisecond-scale lifetimes of modern 3D superconducting cavities \cite{Reagor2016, Milul2023}. By contrast, one could bypass amplification and rely solely on unamplified QPE, where each probabilistic shot is short ($T_{\hat{\mathcal{A}}} \approx 1.5~\mu\text{s}$). However, for $N=200$, achieving success with high probability requires $\eta \approx 105$ independent repetitions, as presented in Table~\ref{tab:qpe_qaa_summary}. The cumulative operational time to successfully herald the state would then exceed $150~\mu\text{s}$ (even excluding the substantial delays for active reset and readout between shots).  Therefore, QAA efficiently trades a massive execution overhead for a moderate increase in coherent depth, yielding a deterministic preparation that is both faster overall and fully compatible with existing coherence limits.

Although QAA effectively bypasses the macroscopic time overhead of probabilistic heralding, its experimental viability ultimately hinges on preserving the quantum state throughout the extended coherent duration of a single amplified execution. Specifically, the resonator amplitude damping must not erase the algorithmic gain from QAA, enforcing the strict requirement $\kappa T_{\text{tot}} \ll 1$. Since state-of-the-art superconducting cavities routinely exhibit single-photon lifetimes $1/\kappa \sim 0.5$--$2~\text{ms}$ \cite{Reagor2016, Milul2023}, the condition $\kappa T_{\text{tot}} \sim 10^{-2}$ is comfortably satisfied for the durations estimated above. Moreover, using representative circuit-QED parameters, our open-system simulations up to $N=200$ still achieve near-unit post-amplification success with $r_y \le 4$, indicating that the dissipation accumulated over the required coherent depth remains manageable for these mesoscopic targets. Crucially, because the algorithmic depth and register size grow exceptionally slowly with the target photon number, the decoherence time overhead remains strictly manageable. This favorable scaling ensures that even for substantially larger macroscopic targets, the preparation protocol avoids exhausting the hardware intrinsic coherence limits, preserving the quantum resource for subsequent applications.

\section{Conclusion and Outlook}\label{sec:conclusion}

We have presented a protocol that converts quantum phase estimation from a passive readout tool into an active preparation scheme for Fock states. The QPE stage encodes photon-number information into a multi-qubit register, while quantum amplitude amplification coherently enhances a marked register outcome before the final measurement. Starting from a coherent field, this procedure overcomes the small Poissonian weight of a prescribed large-$N$ component and enables near-deterministic heralded preparation. For coherent inputs, QAA reduces the resource scaling associated with repeated probabilistic preparation from $\mathcal{O}(1/a_0)$ to $\mathcal{O}(1/\sqrt{a_0})$, at the cost
of a controlled increase in coherent circuit depth, while the required register size grows only logarithmically with the accessible photon-number range. Although our analysis focused on a dispersive
circuit-QED implementation with tunable qubit detunings, the same principle can be adapted to other cavity-QED platforms providing controllable QND photon-number encoding.

The protocol also provides a route for generating two-mode path-entangled states. Directly amplifying the NOON subspace from two coherent inputs is formally possible, but becomes inefficient because its initial probability decreases exponentially with $N$. A more favorable strategy is to first prepare $\ket{N}$ in one resonator and then entangle it with a second resonator initially in the vacuum. This can be implemented by preparing an auxiliary qubit in a superposition, applying a qubit-controlled beam-splitter exchange between the two modes, and measuring the qubit in a rotated basis to erase which-mode information. Such an extension may be connected to parametrically activated beam-splitter interactions between superconducting resonators~\cite{Lu2023HighFidelityBeamsplitter} and to
transmon- or qutrit-mediated protocols for NOON-state generation in circuit-QED~\cite{Su2014NOONCircuitQED}.

Beyond state preparation, well-defined mesoscopic Fock states provide controlled initial conditions for QND monitoring and phase-space tomography. Repeated nondestructive measurements have enabled the observation of progressive field-state collapse, photon-number quantum jumps, and cavity-relaxation trajectories~\cite{Guerlin2007, Quantum_HSJ}. In a companion manuscript currently in preparation, part
of the dispersive QPE architecture developed here is used without amplitude amplification as a number-resolved QND readout for tracking such dissipative trajectories. The same readout can be combined with
phase-space displacements to sample the displaced photon-number statistics entering Wigner-function reconstruction, while the
large-$N$ states generated by the present protocol provide demanding calibration targets for these measurements~\cite{PhysRevLett.78.2547,
Bertet2002, Delglise2008}. 

These capabilities make the protocol relevant to quantum-enhanced metrology~\cite{Giovannetti2011, Deng2024}, bosonic error correction~\cite{PhysRevX.6.031006}, and oscillator-based quantum information processing~\cite{Eickbusch2022}. When combined with
conditional two-mode interactions, the prepared states can also seed NOON states for quantum-enhanced interferometry and related applications~\cite{Boto2000QuantumLithography,
Dowling2008HighNOON, Huver2008}. By combining QND photon-number encoding with coherent probability amplification, this work establishes a scalable route toward the preparation and subsequent manipulation of large nonclassical bosonic resource states.

\section{ACKNOWLEDGMENTS}

The authors thank Luiz Otavio Ribeiro Solak, Alan Costa dos Santos and Juan José García-Ripoll for valuable discussions and suggestions on this work. This study was financed, in part, by the São Paulo Research Foundation (FAPESP), Brazil, Process Number \#2022/00209-6, \#2025/15490-0, \#2024/02604-5, \#2022/10218-2 and \#2025/23694-5, by the Coordenação de Aperfeiçoamento de Pessoal de Nível Superior (CAPES) Finance Code 001, and by the Brazilian National Council for Scientific and Technological Development -- CNPq, Grants No.~140001/2023-9, No.~405712/2023-5, No.~311612/2021-0, and No. 302234/2026-8.

\appendix

\section{Gate-level implementation of QPE and QAA}
\label{app:gates}

\begin{figure*}[!htb]
\includegraphics[width=1.0\textwidth, clip]{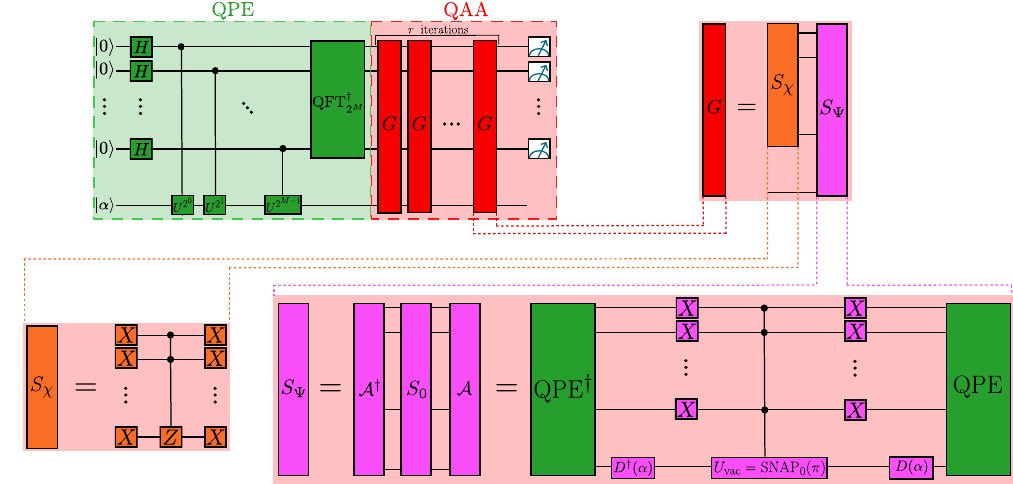}
\caption{Gate-level decomposition of the protocol. 
\emph{Top:} high-level circuit diagram. The QPE stage (green box) prepares a uniform superposition in the $M$-qubit register (Hadamard gates), applies the controlled powers $\hat U_{\rm QPE}^{2^{k-1}}$ generated by dispersive phase accumulation, and concludes with the inverse quantum Fourier transform ($\mathrm{QFT}^\dagger$) to map the photon-number-dependent phase onto a computational outcome $y$. The QAA stage (red box) then applies $r$ Grover iterations (loop) before a final projective measurement of the register.
\emph{Bottom:} decomposition of a single Grover iterate $\hat G=\hat S_{\Psi}\hat S_{\chi}$. The oracle reflection $\hat S_{\chi}=\hat I_r-2\hat \Pi_{\rm good}$ flips the phase of the marked register outcome(s) $y\in{\rm good}$. The diffusion reflection is implemented in the standard amplitude amplification form $\hat S_{\Psi}=\hat{\mathcal{A}}\hat S_{0}\hat{\mathcal{A}}^{\dagger}$, where $\hat{\mathcal{A}}$ denotes the coherent QPE state-preparation map (including the displacement that prepares the initial cavity state) applied prior to the register measurement. Here $\hat S_{0}$ is a reflection about the initialized register state together with a cavity-selective vacuum phase $\hat U_{\rm vac}=\hat I_f-2\ket{0}\!\bra{0}$. Consequently, each Grover iteration reuses the QPE block and its inverse (uncomputation), making explicit that QAA entails repeated QPE applications within the amplification loop.}
\label{fig:scheme}
\end{figure*}

This section focuses on detailing the logical gates underlying the two algorithms used in the main text — Quantum Phase Estimation (QPE)~\cite{kitaev_1995} and Quantum Amplitude Amplification (QAA)~\cite{Brassard2002} — and provides a gate-level decomposition of the full protocol shown in Fig.~\ref{fig:scheme}. In particular, the green (QPE) and red (QAA) blocks in Fig.~\ref{fig:scheme} make explicit how the phase register is prepared, how the photon-number information is coherently encoded into register phases, and how the subsequent Grover iterations reuse the QPE mapping and its inverse within the amplification loop.

\subsection{QPE block: state preparation and phase readout}

The gate sequence to implement the QPE mapping (green box in Fig.~\ref{fig:scheme}) starts from an initial state of Eq.~\eqref{eq:initial_state}, in which the $M$-qubit register is prepared in the computational state $\ket{0}^{\otimes M}$ and the bosonic mode is prepared in the coherent state $\ket{\alpha}$. Hadamard gates are then applied to all register qubits, creating a uniform superposition over computational basis states,
\begin{equation}
H^{\otimes M}\ket{0}^{\otimes M}=\frac{1}{\sqrt{2^M}}\sum_{x=0}^{2^M-1}\ket{x},
\end{equation}
where $x$ labels the $M$-bit computational basis state of the register, i.e., $\ket{x}\equiv\ket{b_{M-1}\cdots b_0}$ with $x=\sum_{j=0}^{M-1} b_j 2^j$ and $b_j\in\{0,1\}$.

Next, the circuit applies the standard controlled powers $\hat U_{\rm QPE}^{2^{k-1}}$ (with $k=1,\ldots,M$) of Eq.~\eqref{eq:U_qpe}. A detailed derivation of this photon-number unitary from the dispersive circuit-QED interaction, together with the corresponding calibration condition, is provided in  Appendix~\ref{app:uqpe_from_hint}. It is equivalently described as the application of $\hat U_{\rm QPE}^{x}$ conditioned on the register integer $x$. Acting on an arbitrary field state $\ket{\psi_f}$, the joint state after the controlled-$\hat U_{\rm QPE}$ stage can be written as
\begin{equation}
\ket{\Psi_0}=\frac{1}{\sqrt{2^M}}\sum_{x=0}^{2^M-1}\ket{x}\otimes \hat U_{\rm QPE}^{x}\ket{\psi_f}.
\label{eq:psi_ctrl}
\end{equation}

For an eigenstate $\ket{n}$ of $\hat n$, one has $\hat U_{\rm QPE}^{x}\ket{n}=\exp(-i 2\pi n x/2^M)\ket{n}$, so that the information about $n$ appears as a linear phase gradient across the register basis states $\ket{x}$. The inverse quantum Fourier transform $\mathrm{QFT}^\dagger$ is then applied to the register, converting this phase gradient into a distribution that is sharply localized in the computational basis~\cite{cleve1998quantum, Torosov2009, nielsen2010quantum}. Concretely, $\mathrm{QFT}^\dagger$ performs the discrete Fourier decoding that maps the phase-encoded superposition onto a binary estimate of the eigenphase. To avoid confusion in notation, we denote by $x$ the register integer labeling the computational basis before $\mathrm{QFT}^\dagger$ (i.e., the control index of the powers $\hat U_{\rm QPE}^{x}$), whereas $y$ denotes the computational basis outcome after $\mathrm{QFT}^\dagger$, which provides the binary estimate of the encoded eigenphase. Finally, we emphasize that in our protocol the QPE mapping is applied without an intermediate measurement, since the subsequent QAA step requires access to the joint QPE output state for uncomputation within each Grover iteration.

\subsection{QAA block: oracle and diffusion reflections}

We now detail the gate decomposition of the QAA stage (red box in Fig.~\ref{fig:scheme}). In our protocol, amplitude amplification acts on the unmeasured joint output of the QPE block. 
Specifically, we define
\begin{equation}
\ket{\Psi_0}\equiv \hat{\mathcal{A}}\ket{0}^{\otimes M}\ket{\alpha},
\end{equation}
where $\hat{\mathcal{A}}$ denotes the coherent state-preparation map prior to measurement, i.e., the same operations used to prepare the QPE output state (displacement preparing $\ket{\alpha}$ together with the QPE mapping on the register).
QAA then boosts the probability of obtaining the marked register outcomes by repeatedly applying the Grover iterate $\hat G=\hat S_{\Psi}\hat S_{\chi}$~\cite{Brassard2002}, where $\hat S_{\chi}$ is the oracle (marking) reflection on the register and $\hat S_{\Psi}$ is the diffusion (state) reflection about $\ket{\Psi_0}$.

\subsubsection{Oracle reflection $\hat S_{\chi}$}

Let ${\rm good}\subset\{0,\dots,N_\text{max}-1\}$ be the accepted set of register outcomes (often a single $y_0$). 
With the projector $\hat \Pi_{\rm good}$ defined in Eq.~\eqref{eq:proj_good}, the oracle reflection is $S_{\chi}=\hat I_r-2\hat \Pi_{\rm good}$, where $\hat I_r$ denotes the identity operator on the register Hilbert space. For a single marked outcome $y_0$ written in binary as $y_0\equiv b_{M-1}\cdots b_0$ with $b_k\in\{0,1\}$, a standard compilation is illustrated in Fig.~\ref{fig:scheme}: one applies $X$ gates on every qubit $k$ such that $b_k=0$, thereby mapping $\ket{y_0}$ to $\ket{11\cdots 1}$ (since $X$ flips $\ket{0}\leftrightarrow\ket{1}$), then applies an $M$-controlled phase flip (multi-controlled $Z$) that adds a phase $-1$ ($\pi$) only to $\ket{11\cdots 1}$, and finally applies the same $X$ gates again to undo the mapping. 
This implements a phase flip on $\ket{y_0}$ and leaves all other computational basis states unchanged. 
Efficient decompositions of multi-controlled phase gates into one- and two-qubit primitives are well known~\cite{Barenco1995, Glaser2023, Chen2024}.

\subsubsection{Diffusion reflection $\hat S_{\Psi}$}

The diffusion step is implemented in the standard amplitude amplification form $\hat S_{\Psi}=\hat{\mathcal{A}}\hat S_0 \hat{\mathcal{A}}^{\dagger}$~\cite{Brassard2002}, where $\hat S_0$ is a reflection about the initialized reference state. 
In our setting, $S_0$ factors into a register part (a phase flip conditioned on $\ket{0}^{\otimes M}$) and a cavity part that applies a vacuum-selective phase, $\hat U_{\rm vac}=\hat I_f-2\ket{0}\!\bra{0}$, where denotes the identity operator on the field Hilbert space. In circuit QED, such number-selective phases are naturally implemented via selective number-dependent arbitrary phase (SNAP) operations~\cite{Schuster2007, Heeres2015, Krastanov2015}. Moreover, a reflection about the coherent component can be expressed by conjugation with displacements, $\hat D(\alpha)\hat U_{\rm vac}\hat D^{\dagger}(\alpha)$, consistent with the state preparation in $\hat{\mathcal{A}}$.

Because $\hat S_{\Psi}$ explicitly contains $\hat{\mathcal{A}}$ and $\hat{\mathcal{A}}^{\dagger}$, each Grover iteration reuses the QPE preparation and its inverse (uncomputation): beyond the initial coherent QPE preparation of $\ket{\Psi_0}$, the QAA loop contributes one additional application of $\hat{\mathcal{A}}$ and one of $\hat{\mathcal{A}}^{\dagger}$ per iteration, together with the two reflections $\hat S_{\chi}$ and $\hat S_0$. 
This is precisely the sense in which QAA ``repeats QPE'' within the amplification loop, as highlighted in Fig.~\ref{fig:scheme}.

\section{From the dispersive interaction to the QPE unitary}\label{app:uqpe_from_hint}

The main text introduces the photon-number unitary $\hat U_{\rm QPE}$ in Eq.~\eqref{eq:U_qpe} as the eigenphase map targeted by quantum phase estimation (QPE). The goal of this section is to make explicit that this unitary is not an arbitrary choice, but rather a natural logical abstraction of the physical dispersive evolution in circuit-QED. This direct correspondence yields the calibration condition used to implement the controlled powers $\hat U_{\rm QPE}^{2^{k-1}}$.

\subsection{Effective dispersive Hamiltonian from a Schrieffer--Wolff transformation}

After the resonator being prepared in a coherent state by a classical pump (which is subsequently turned off), the coherent dynamics of the system are governed by the multiqubit Jaynes--Cummings Hamiltonian~\cite{JaynesCummings1963} ($\hbar=1$),
\begin{align}
\hat H&=\omega_c\,\hat a^\dagger\hat a+\sum_{k=1}^{M}\frac{\omega_{q_k}}{2}\,\hat\sigma_z^{(k)} \nonumber \\
&\quad +\sum_{k=1}^{M} g_k\left(\hat a\,\hat\sigma_+^{(k)}+\hat a^\dagger \hat\sigma_-^{(k)}\right),
\label{eq:HJC_multi}
\end{align}
where $\hat a$ ($\hat a^\dagger$) annihilates (creates) photons in the resonator, $\omega_{q_k}$ is the transition frequency of the $k$-th qubit, and $\hat\sigma_-^{(k)}$ ($\hat\sigma_+^{(k)}$) is the atomic lowering (raising) operator for the $k$-th qubit.

 In the dispersive regime, characterized by large detunings
\begin{equation}
|\Delta_k|\equiv|\omega_{q_k}-\omega_c|\gg g_k\sqrt{\bar n+1},
\end{equation}
where $\bar n=\langle \hat n\rangle$ is the intracavity photon number, the exchange terms oscillate rapidly. The qubit--resonator coupling can then be diagonalized perturbatively in the small parameter $g_k/\Delta_k$ via a Schrieffer--Wolff (``small-rotation'') transformation~\cite{JaynesCummings1963,Schuster2007,BlaisRMP2021}. Concretely, applying a unitary $\hat U=\exp(\hat S)$ with the anti-Hermitian generator
\begin{equation}
\hat S=\sum_{k=1}^{M}\frac{g_k}{\Delta_k}\left(\hat a\,\hat\sigma_+^{(k)}-\hat a^\dagger\hat\sigma_-^{(k)}\right),
\end{equation}
and expanding $\hat U\hat H\hat U^\dagger$ using the Baker--Campbell--Hausdorff series,
\begin{equation}
\hat U\hat H\hat U^\dagger
=\hat H+[\hat S,\hat H]+\frac{1}{2}[\hat S,[\hat S,\hat H]]+\cdots,
\end{equation}
truncating at the second order in $g_k/\Delta_k$. This truncation is controlled because the dispersive condition ensures $g_k/\Delta_k\ll 1$ over the relevant photon-number manifold.

A dispersive coupling is obtained between the photon number operator $\hat n\equiv\hat a^\dagger\hat a$ and the atomic inversion operator $\hat \sigma_z^{(k)}$. Moving to the interaction picture with respect to the renormalized free Hamiltonian (e.g., $\hat H_0=\omega_c \hat n+\sum_k \tilde\omega_{q_k}\hat\sigma_z^{(k)}/2$), which absorbs Lamb and Stark shifts), the relevant entangling evolution is generated by
\begin{equation}
\hat H_{\rm int}=\sum_{k=1}^{M}\chi_k\,\hat n\,\hat\sigma_z^{(k)},
\label{eq:Hint_app}
\end{equation}
with the dispersive shift $\chi_k\simeq g_k^2/\Delta_k$.

\subsection{Dispersive phase accumulation}

The dispersive interaction in Eq.~\eqref{eq:Hint_app} implements a photon-number-dependent phase shift on the multiqubit register. Focusing on a single qubit $k$, the evolution operator in the interaction picture is
\begin{equation}
\hat U_{\rm int}^{(k)}(t)=\exp\!\left(-i\,\chi_k\,\hat n\,\hat\sigma_z^{(k)}\,t\right),
\end{equation}
where this evolution is equivalent to a controlled photon-number phase acting on the field conditioned on the qubit state. For a Fock component $\ket{n}$ and a qubit initialized in the superposition $\ket{+}_k=(\ket{0_k}+\ket{1_k})/\sqrt{2}$, the joint state evolves as
\begin{equation}
\ket{+}_k\ket{n}\ \longrightarrow\ \frac{\ket{0_k}+e^{-i\,\chi_k\,n\,t}\ket{1_k}}{\sqrt{2}}\ket{n}.
\label{eq:phase_accum_app}
\end{equation}

Equation~\eqref{eq:phase_accum_app} reveals that the physical dispersive interaction inherently performs the exact phase-encoding primitive required by QPE. Because the accumulated phase is strictly proportional to the photon number $n$, the logical operation we are applying to the field is, by definition, the unitary
\begin{equation}
\hat U_{\rm QPE}=\exp\!\left(-i\,\frac{2\pi}{N_\text{max}}\,\hat n\right),
\label{eq:Uqpe_app}
\end{equation}
where $N_\text{max}$ represents the maximum number of distinguishable phases (with $N_\text{max}=2^M$). Defining $\hat U_{\rm QPE}$ in this way is therefore not an arbitrary algorithmic choice, but the direct mathematical representation of the normalized dispersive phase. The eigenstates are precisely the Fock states $\ket{n}$, and the encoded eigenphases $2\pi n/N_\text{max}$ correspond directly to the binary fraction $\theta_y=y/N_\text{max}$ estimated by the QPE register outcome $y\in\{0,\ldots,N_\text{max}-1\}$.

\subsection{Calibration condition for the controlled powers $\hat U_{\rm QPE}^{2^{k-1}}$}

In the standard formulation of QPE, the $k$-th register qubit controls the power $\hat U_{\rm QPE}^{2^{k-1}}$. Acting on a Fock component $\ket{n}$, the target algorithmic phase is
\begin{equation}
\hat U_{\rm QPE}^{2^{k-1}}\ket{n} = \exp\left(-i\,\frac{2\pi\,2^{k-1}}{N_\text{max}}\,n\right)\ket{n}.
\end{equation}
To implement this target unitary physically using the dispersive evolution over a fixed interaction time $t_{\rm int}$, the accumulated physical phase must identically match the algorithmic phase:
\begin{align}
\exp\left(-i\chi_k\, n\,t_\text{int} \right) &= \exp\left(-i\,\frac{2\pi\,2^{k-1}}{N_\text{max}}\,n\right) \\
\chi_k\,t_{\rm int}&=\frac{2\pi\,2^{k-1}}{N_\text{max}}.
\label{eq:chi_calib_app}
\end{align}

\section{NOON state generation}\label{app:noon_extension}

This appendix discusses how the large Fock-state resource prepared by the protocol could be converted into a two-mode NOON state and why a direct two-mode amplification strategy is not scalable. We define
\begin{equation}
    \ket{\mathrm{NOON}}_{N,\phi}
    =
    \frac{1}{\sqrt{2}}
    \left(
    \ket{N}_a\ket{0}_b
    +
    e^{i\phi}\ket{0}_a\ket{N}_b
    \right),
    \label{eq:noon_definition_app}
\end{equation}
where all $N$ photons coherently occupy either mode $a$ or mode $b$. Such states are widely discussed as resources for quantum-enhanced metrology and related interferometric applications~\cite{Boto2000QuantumLithography,Dowling2008HighNOON,Huver2008,Giovannetti2011}.

\subsection{Direct two-mode amplification of the NOON subspace}

A direct extension of the protocol would start from two coherent modes,
\begin{equation}
    \ket{\psi}_{\rm field}=\ket{\alpha}_a\ket{\beta}_b,
\end{equation}
and use QPE to encode the two-mode photon-number pair $(n_a,n_b)$ into auxiliary register degrees of freedom. The QAA oracle would then coherently mark the subspace
\begin{equation}
    \mathcal{G}_N=\mathrm{span}\{\ket{N,0}_{ab},\ket{0,N}_{ab}\}.
\end{equation}
The relevant component of the initial two-mode coherent state is
\begin{equation}
    \Pi_{\mathcal{G}_N}\ket{\alpha}_a\ket{\beta}_b
    =
    e^{-(|\alpha|^2+|\beta|^2)/2}
    \frac{\alpha^N\ket{N,0}_{ab}+\beta^N\ket{0,N}_{ab}}{\sqrt{N!}}.
\end{equation}
For $|\alpha|=|\beta|$ the two branches have equal weight, and choosing their phases such that $\beta^N/\alpha^N=e^{i\phi}$ gives the normalized state in Eq.~\eqref{eq:noon_definition_app}. Thus, the direct strategy is formally capable of producing a NOON state if the marking operation preserves coherence between the two branches and does not reveal which of them occurred.

The limitation is the initial probability of the marked subspace. For a single-mode Fock target prepared from a coherent state with $|\alpha|^2=N$, the initial target probability is presented in Eq.~\eqref{eq:a0_QPE} with the optimal iteration defined in Eq.~\eqref{eq:ry}, which gives $r^{\rm Fock}\propto N^{1/4}$ for coherent inputs. For the two-mode NOON subspace, the marked probability is
\begin{equation}
    a_0^{\rm NOON}
    =
    e^{-(|\alpha|^2+|\beta|^2)}
    \frac{|\alpha|^{2N}+|\beta|^{2N}}{N!}.
    \label{eq:a0_noon_general_app}
\end{equation}
A balanced NOON state requires $|\alpha|=|\beta|$. Writing $|\alpha|^2=|\beta|^2=\mu$, Eq.~\eqref{eq:a0_noon_general_app} becomes
\begin{equation}
    a_0^{\rm NOON}(\mu)=2e^{-2\mu}\frac{\mu^N}{N!}.
\end{equation}
Maximizing this expression with respect to $\mu$ gives $\mu=N/2$, so the optimal balanced coherent input has total mean photon number $N$. At this optimum,
\begin{equation}
    a_0^{\rm NOON}
    =
    2e^{-N}\frac{(N/2)^N}{N!}
    \simeq
    \frac{2^{1-N}}{\sqrt{2\pi N}}.
    \label{eq:a0_noon_app}
\end{equation}
Therefore,
\begin{equation}
    a_0^{\rm NOON}=2^{1-N}a_0^{\rm Fock},
\end{equation}
and yields
\begin{equation}
    r^{\rm NOON}
    \simeq
    \frac{\pi}{4}
    2^{(N-1)/2}(2\pi N)^{1/4}
    -
    \frac{1}{2}.
\end{equation}
Thus, the direct two-mode coherent-state strategy scales as
\begin{equation}
    r^{\rm NOON}\propto 2^{N/2}N^{1/4},
\end{equation}
which is exponentially worse than the single-mode Fock-state preparation scaling. This motivates using the presented protocol only to prepare $\ket{N}$ in one mode and performing the two-mode entangling step afterward.

\subsection{Using the prepared Fock state as a NOON state}

Assume that the protocol of the main text has prepared a Fock state in mode $a$, while a second bosonic mode $b$ is initialized in the vacuum,
\begin{equation}
    \ket{\psi}_{\rm field}=\ket{N}_a\ket{0}_b\equiv\ket{N,0}_{ab}.
\end{equation}
We introduce an auxiliary qubit initialized in $\ket{0}_q$, the initial state will be:
\begin{equation}
    \ket{\psi_0} = \ket{0}_q \otimes \ket{N,0}_{ab}.
\end{equation}
Applying a Hadamard-type rotation with a controllable phase, we obtain
\begin{equation}
    \ket{\psi_1} = \frac{\ket{0}_q+e^{i\phi_q}\ket{1}_q}{\sqrt{2}}\otimes \ket{N,0}_{ab}.
\end{equation}
The ideal logical operation is a controlled swap between the two bosonic modes. Such a bosonic controlled-SWAP (Fredkin) operation has been experimentally implemented in circuit-QED using two microwave cavity modes and a transmon ancilla~\cite{Gao2019}. At the logical
level, it is written as
\begin{equation}
    \hat U_{\rm control-S}
    =
    \ket{0}_q\bra{0}\otimes\hat I_{ab}
    +
    \ket{1}_q\bra{1}\otimes\hat S_{ab},
\end{equation}
where $\hat S_{ab}\ket{n_a,n_b}_{ab}=\ket{n_b,n_a}_{ab}$. It maps the state to
\begin{equation}
    \ket{\psi_2}
    =
    \frac{1}{\sqrt{2}}
    \left(
    \ket{0}_q\ket{N,0}_{ab}
    +
    e^{i\phi_q}\ket{1}_q\ket{0,N}_{ab}
    \right).
    \label{eq:ideal_cswap_state_app}
\end{equation}
The qubit stores the mode in which the information is. Measuring it in the computational basis would therefore collapse the field onto either $\ket{N,0}_{ab}$ or $\ket{0,N}_{ab}$. Instead, measuring the qubit in the rotated basis $\ket{\pm}_q=(\ket{0}_q\pm\ket{1}_q)/\sqrt{2}$ erases this information. Rewriting Eq.~\eqref{eq:ideal_cswap_state_app} in that basis gives
\begin{align}
    \ket{\psi_2}
    &=
    \frac{1}{2}\ket{+}_q
    \left(
    \ket{N,0}_{ab}+e^{i\phi_q}\ket{0,N}_{ab}
    \right)
    \nonumber\\
    &\quad+
    \frac{1}{2}\ket{-}_q
    \left(
    \ket{N,0}_{ab}-e^{i\phi_q}\ket{0,N}_{ab}
    \right).
\end{align}
Thus, both measurement outcomes herald NOON states, differing only by a known relative phase.

\subsection{Beam-splitter realization of the mode swap}

The physical operation underlying the mode swap can be generated by a beam-splitter Hamiltonian,
\begin{equation}
    \hat H_{\rm BS}(t)
    =
    \hbar g_{\rm BS}(t)
    \left(
    e^{i\varphi_{\rm BS}}\hat a^{\dagger}\hat b
    +
    e^{-i\varphi_{\rm BS}}\hat a\hat b^{\dagger}
    \right),
    \label{eq:H_bs_app}
\end{equation}
where $g_{\rm BS}(t)$ is the exchange rate and $\varphi_{\rm BS}$ is a controllable drive phase. Such bilinear interactions can be engineered parametrically between microwave resonators using nonlinear couplers and external drives~\cite{Lu2023HighFidelityBeamsplitter}. More generally, effective beam-splitter interactions between two cavity modes can also be mediated by driven two- or three-level atoms, including state-dependent couplings that provide a natural route toward conditional mode exchange~\cite{VillasBoas2005, Prado2006}. Defining
\begin{align}
    \hat K &=
    e^{i\varphi_{\rm BS}}\hat a^{\dagger}\hat b
    +
    e^{-i\varphi_{\rm BS}}\hat a\hat b^{\dagger}, \\
    \Theta &=\int_0^{t_{\rm BS}}g_{\rm BS}(t')dt',
\end{align}
the unitary is $\hat U_{\rm BS}=\exp(-i\Theta\hat K)$ if $\varphi_{\rm BS}$ is constant during the pulse. The creation mode operator transformation is
\begin{equation}
    \hat U_{\rm BS}\hat a^{\dagger}\hat U_{\rm BS}^{\dagger}
    =
    \hat a^{\dagger}\cos\Theta
    -
    i e^{-i\varphi_{\rm BS}}\hat b^{\dagger}\sin\Theta,
    \label{eq:creation_transform_app}
\end{equation}
which follows directly from the Baker--Campbell--Hausdorff expansion using $[\hat K,\hat a^{\dagger}]=e^{-i\varphi_{\rm BS}}\hat b^{\dagger}$ and $[\hat K,\hat b^{\dagger}]=e^{i\varphi_{\rm BS}}\hat a^{\dagger}$.
Since $\ket{N,0}_{ab}=(\hat a^{\dagger})^N\ket{0,0}_{ab}/\sqrt{N!}$,
\begin{equation}
    \hat U_{\rm BS}\ket{N,0}_{ab}
    =
    \frac{\left(\hat a^{\dagger}\cos\Theta-i e^{-i\varphi_{\rm BS}}\hat b^{\dagger}\sin\Theta\right)^N}{\sqrt{N!}}
    \ket{0,0}_{ab}.
\end{equation}
For a complete swap, $\Theta=\pi/2$, and therefore
\begin{equation}
    \hat U_{\rm BS}\left(\frac{\pi}{2}\right)\ket{N,0}_{ab}
    =
    e^{i\varphi_{\rm swap}(N)}\ket{0,N}_{ab},
    \label{eq:swap_phase_app}
\end{equation}
with
\begin{equation}
    e^{i\varphi_{\rm swap}(N)}=e^{-iN(\varphi_{\rm BS}+\pi/2)}.
\end{equation}
The phase is deterministic and can be incorporated into the target NOON phase.

The conditional version is obtained at the logical level by making Eq.~\eqref{eq:H_bs_app} dependent on the auxiliary qubit state,
\begin{equation}
    \hat H_{\rm control-BS}(t)=\ket{1}_q\bra{1}\otimes \hat H_{\rm BS}(t),
    \label{eq:H_cbs_app}
\end{equation}
so that
\begin{equation}
    \hat U_{\rm control-BS}
    =
    \ket{0}_q\bra{0}\otimes \hat I_{ab}
    +
    \ket{1}_q\bra{1}\otimes \hat U_{\rm BS}.
\end{equation}
For $\Theta=\pi/2$, acting on the prepared state gives
\begin{equation}
    \ket{\psi_2}
    =
    \frac{1}{\sqrt{2}}
    \left[
    \ket{0}_q\ket{N,0}_{ab}
    +
    e^{i\Phi}\ket{1}_q\ket{0,N}_{ab}
    \right],
    \label{eq:state_before_measure_app}
\end{equation}
where $\Phi=\phi_q+\varphi_{\rm swap}(N)$ may also include additional deterministic dynamical phases from a concrete implementation.

One possible circuit-QED implementation is to activate the parametric beam-splitter interaction between two resonators through a nonlinear Josephson coupler~\cite{Lu2023HighFidelityBeamsplitter}, while making the exchange conditional on the auxiliary-qubit state. At the effective level, this can be achieved by making the conversion
resonance qubit-state dependent through dispersive shifts. If the qubit shifts the effective mode frequencies as
$\omega_a^{(s)}=\omega_a+\chi_a^{(s)}$ and
$\omega_b^{(s)}=\omega_b+\chi_b^{(s)}$, with $s=0,1$, the conversion drive can be chosen to be resonant for one qubit branch and
off-resonant for the other, approximating
Eq.~\eqref{eq:H_cbs_app}. Alternatively, the same logical controlled-SWAP can be compiled from parametrically driven beam-splitter operations and ancilla-conditioned phase operations, as demonstrated experimentally for two superconducting microwave
cavities~\cite{Gao2019}. Transmon- or qutrit-mediated resonator interactions provide another route, including protocols specifically
designed for NOON-state generation~\cite{Su2014NOONCircuitQED}.

Finally, measuring the auxiliary qubit in the $\{\ket{+},\ket{-}\}$ basis can be implemented using the usual computational-basis readout preceded by a Hadamard gate, since $H\ket{+}=\ket{0}$ and $H\ket{-}=\ket{1}$. If the $\ket{-}$ outcome is obtained, the heralded state has phase $\Phi+\pi$. This relative sign can either be tracked classically or corrected by a single-mode phase rotation on mode $b$,
\begin{equation}
    \hat R_b(\theta)=e^{i\theta\hat b^{\dagger}\hat b},
\end{equation}
which acts as $\hat R_b(\theta)\ket{0,N}_{ab}=e^{iN\theta}\ket{0,N}_{ab}$. Choosing $\theta=\pi/N$ changes the sign of the $\ket{0,N}_{ab}$ component and converts the $\ket{-}$ heralded state into the same NOON phase obtained from the $\ket{+}$ outcome.

\bibliography{ref}

@article{Deng2024,
  title = {Quantum-enhanced metrology with large {Fock} states},
  volume = {20},
  ISSN = {1745-2481},
  url = {http://dx.doi.org/10.1038/s41567-024-02619-5},
  DOI = {10.1038/s41567-024-02619-5},
  number = {12},
  journal = {Nat. Phys.},
  publisher = {Springer Science and Business Media LLC},
  author = {Deng,  Xiaowei and Li,  Sai and Chen,  Zi-Jie and Ni,  Zhongchu and Cai,  Yanyan and Mai,  Jiasheng and Zhang,  Libo and Zheng,  Pan and Yu,  Haifeng and Zou,  Chang-Ling and Liu,  Song and Yan,  Fei and Xu,  Yuan and Yu,  Dapeng},
  year = {2024},
  month = aug,
  pages = {1874}
}

@article{PhysRevLett.76.1796,
  title = {Generation of Nonclassical Motional States of a Trapped Atom},
  author = {Meekhof, D. M. and Monroe, C. and King, B. E. and Itano, W. M. and Wineland, D. J.},
  journal = {Phys. Rev. Lett.},
  volume = {76},
  issue = {11},
  pages = {1796},
  numpages = {0},
  year = {1996},
  month = {Mar},
  publisher = {American Physical Society},
  doi = {10.1103/PhysRevLett.76.1796},
  url = {https://link.aps.org/doi/10.1103/PhysRevLett.76.1796}
}

@article{PhysRevLett.88.143601,
  title = {Generating and Probing a Two-Photon {Fock} State with a Single Atom in a Cavity},
  author = {Bertet, P. and Osnaghi, S. and Milman, P. and Auffeves, A. and Maioli, P. and Brune, M. and Raimond, J. M. and Haroche, S.},
  journal = {Phys. Rev. Lett.},
  volume = {88},
  issue = {14},
  pages = {143601},
  numpages = {4},
  year = {2002},
  month = {Mar},
  publisher = {American Physical Society},
  doi = {10.1103/PhysRevLett.88.143601},
  url = {https://link.aps.org/doi/10.1103/PhysRevLett.88.143601}
}

@article{Varcoe2000,
  title = {Preparing pure photon number states of the radiation field},
  volume = {403},
  ISSN = {1476-4687},
  url = {http://dx.doi.org/10.1038/35001526},
  DOI = {10.1038/35001526},
  number = {6771},
  journal = {Nature (London)},
  publisher = {Springer Science and Business Media LLC},
  author = {Varcoe,  B. T. H. and Brattke,  S. and Weidinger,  M. and Walther,  H.},
  year = {2000},
  month = feb,
  pages = {743}
}

@article{Guerlin2007,
  title = {Progressive field-state collapse and quantum non-demolition photon counting},
  volume = {448},
  ISSN = {1476-4687},
  url = {http://dx.doi.org/10.1038/nature06057},
  DOI = {10.1038/nature06057},
  number = {7156},
  journal = {Nature (London)},
  publisher = {Springer Science and Business Media LLC},
  author = {Guerlin,  Christine and Bernu,  Julien and Deléglise,  Samuel and Sayrin,  Clément and Gleyzes,  Sébastien and Kuhr,  Stefan and Brune,  Michel and Raimond,  Jean-Michel and Haroche,  Serge},
  year = {2007},
  month = aug,
  pages = {889}
}

@article{PhysRevLett.125.093603,
  title = {Deterministic Generation of Large {Fock} States},
  author = {Uria, M. and Solano, P. and Hermann-Avigliano, C.},
  journal = {Phys. Rev. Lett.},
  volume = {125},
  issue = {9},
  pages = {093603},
  numpages = {6},
  year = {2020},
  month = {Aug},
  publisher = {American Physical Society},
  doi = {10.1103/PhysRevLett.125.093603},
  url = {https://link.aps.org/doi/10.1103/PhysRevLett.125.093603}
}

@article{Liu2004,
  title = {Generation of nonclassical photon states using a superconducting qubit in a microcavity},
  volume = {67},
  ISSN = {1286-4854},
  url = {http://dx.doi.org/10.1209/epl/i2004-10144-3},
  DOI = {10.1209/epl/i2004-10144-3},
  number = {6},
  journal = {Europhys. Lett.},
  publisher = {IOP Publishing},
  author = {Liu,  Yu-xi and Wei,  L. F and Nori,  Franco},
  year = {2004},
  month = sep,
  pages = {941}
}

@article{Hofheinz2008,
  title = {Generation of {Fock} states in a superconducting quantum circuit},
  volume = {454},
  ISSN = {1476-4687},
  url = {http://dx.doi.org/10.1038/nature07136},
  DOI = {10.1038/nature07136},
  number = {7202},
  journal = {Nature (London)},
  publisher = {Springer Science and Business Media LLC},
  author = {Hofheinz,  Max and Weig,  E. M. and Ansmann,  M. and Bialczak,  Radoslaw C. and Lucero,  Erik and Neeley,  M. and O’Connell,  A. D. and Wang,  H. and Martinis,  John M. and Cleland,  A. N.},
  year = {2008},
  month = jul,
  pages = {310}
}

@article{McCormick2019,
  title = {Quantum-enhanced sensing of a single-ion mechanical oscillator},
  author = {McCormick, Katherine C. and Keller, Jonas and Burd, Shaun C. and Wineland, David J. and Wilson, Andrew C. and Leibfried, Dietrich},
  journal={Nature (London)},
  volume={572},
  number={7767},
  pages={86},
  year={2019},
  publisher={Nature Publishing Group UK London},
  doi = {10.1038/s41586-019-1421-y},
  url = {https://doi.org/10.1038/s41586-019-1421-y}
}

@article{Meher2026,
  title = {Generation of large Fock states from coherent states using Kerr interaction and displacement},
  author = {Meher, Nilakantha and Pathak, Anirban and Sivakumar, S.},
  journal = {Phys. Rev. A},
  volume = {113},
  issue = {3},
  pages = {033704},
  numpages = {8},
  year = {2026},
  month = {Mar},
  publisher = {American Physical Society},
  doi = {10.1103/6x7s-5696},
  url = {https://link.aps.org/doi/10.1103/6x7s-5696}
}

@article{Xiong2026,
  title = {Scalable high-fidelity and near-deterministic preparation of large-photon-number states},
  author = {Xiong, Mo and Han, Jize and Cao, Chuanzhen and Li, Jinbin and Huang, Zhiguo and Xue, Ming},
  journal = {Phys. Rev. A},
  volume = {114},
  issue = {1},
  pages = {013721},
  numpages = {10},
  year = {2026},
  month = {Jul},
  publisher = {American Physical Society},
  doi = {10.1103/f49z-hqgh},
  url = {https://link.aps.org/doi/10.1103/f49z-hqgh}
}

@article{Wang2017,
  title = {Converting Quasiclassical States into Arbitrary {Fock} State Superpositions in a Superconducting Circuit},
  author = {Wang, W. and Hu, L. and Xu, Y. and Liu, K. and Ma, Y. and Zheng, Shi-Biao and Vijay, R. and Song, Y. P. and Duan, L.-M. and Sun, L.},
  journal = {Phys. Rev. Lett.},
  volume = {118},
  issue = {22},
  pages = {223604},
  numpages = {6},
  year = {2017},
  month = {Jun},
  publisher = {American Physical Society},
  doi = {10.1103/PhysRevLett.118.223604},
  url = {https://link.aps.org/doi/10.1103/PhysRevLett.118.223604}
}

@article{Chu2018,
  title = {Creation and control of multi-phonon {Fock} states in a bulk acoustic-wave resonator},
  volume = {563},
  ISSN = {1476-4687},
  url = {http://dx.doi.org/10.1038/s41586-018-0717-7},
  DOI = {10.1038/s41586-018-0717-7},
  number = {7733},
  journal = {Nature (London)},
  publisher = {Springer Science and Business Media LLC},
  author = {Chu,  Yiwen and Kharel,  Prashanta and Yoon,  Taekwan and Frunzio,  Luigi and Rakich,  Peter T. and Schoelkopf,  Robert J.},
  year = {2018},
  month = nov,
  pages = {666}
}

@article{Eickbusch2022,
  title = {Fast universal control of an oscillator with weak dispersive coupling to a qubit},
  volume = {18},
  ISSN = {1745-2481},
  url = {http://dx.doi.org/10.1038/s41567-022-01776-9},
  DOI = {10.1038/s41567-022-01776-9},
  number = {12},
  journal = {Nat. Phys.},
  publisher = {Springer Science and Business Media LLC},
  author = {Eickbusch,  Alec and Sivak,  Volodymyr and Ding,  Andy Z. and Elder,  Salvatore S. and Jha,  Shantanu R. and Venkatraman,  Jayameenakshi and Royer,  Baptiste and Girvin,  S. M. and Schoelkopf,  Robert J. and Devoret,  Michel H.},
  year = {2022},
  month = oct,
  pages = {1464}
}

@article{Giovannetti2011,
  title = {Advances in quantum metrology},
  volume = {5},
  ISSN = {1749-4893},
  url = {http://dx.doi.org/10.1038/nphoton.2011.35},
  DOI = {10.1038/nphoton.2011.35},
  number = {4},
  journal = {Nat. Photon.},
  publisher = {Springer Science and Business Media LLC},
  author = {Giovannetti,  Vittorio and Lloyd,  Seth and Maccone,  Lorenzo},
  year = {2011},
  month = mar,
  pages = {222}
}

@article{PhysRevX.6.031006,
  title = {New Class of Quantum Error-Correcting Codes for a Bosonic Mode},
  author = {Michael, Marios H. and Silveri, Matti and Brierley, R. T. and Albert, Victor V. and Salmilehto, Juha and Jiang, Liang and Girvin, S. M.},
  journal = {Phys. Rev. X},
  volume = {6},
  issue = {3},
  pages = {031006},
  numpages = {26},
  year = {2016},
  month = {Jul},
  publisher = {American Physical Society},
  doi = {10.1103/PhysRevX.6.031006},
  url = {https://link.aps.org/doi/10.1103/PhysRevX.6.031006}
}

@article{Hofheinz2009,
  title = {Synthesizing arbitrary quantum states in a superconducting resonator},
  volume = {459},
  ISSN = {1476-4687},
  url = {http://dx.doi.org/10.1038/nature08005},
  DOI = {10.1038/nature08005},
  number = {7246},
  journal = {Nature (London)},
  publisher = {Springer Science and Business Media LLC},
  author = {Hofheinz,  Max and Wang,  H. and Ansmann,  M. and Bialczak,  Radoslaw C. and Lucero,  Erik and Neeley,  M. and O’Connell,  A. D. and Sank,  D. and Wenner,  J. and Martinis,  John M. and Cleland,  A. N.},
  year = {2009},
  month = may,
  pages = {546}
}

@article{Sayrin2011,
  title = {Real-time quantum feedback prepares and stabilizes photon number states},
  volume = {477},
  ISSN = {1476-4687},
  url = {http://dx.doi.org/10.1038/nature10376},
  DOI = {10.1038/nature10376},
  number = {7362},
  journal = {Nature (London)},
  publisher = {Springer Science and Business Media LLC},
  author = {Sayrin,  Clément and Dotsenko,  Igor and Zhou,  Xingxing and Peaudecerf,  Bruno and Rybarczyk,  Théo and Gleyzes,  Sébastien and Rouchon,  Pierre and Mirrahimi,  Mazyar and Amini,  Hadis and Brune,  Michel and Raimond,  Jean-Michel and Haroche,  Serge},
  year = {2011},
  month = aug,
  pages = {73}
}

@misc{kitaev_1995,
  doi = {10.48550/ARXIV.QUANT-PH/9511026},
  url = {https://arxiv.org/abs/quant-ph/9511026},
  author = {Kitaev,  A. Yu.},
  title = {Quantum measurements and the {Abelian Stabilizer Problem}},
  publisher = {arXiv},
  year = {1995},
  copyright = {Assumed arXiv.org perpetual,  non-exclusive license to distribute this article for submissions made before January 2004}
}

@book{nielsen2010quantum,
  title={Quantum Computation and Quantum Information},
  author={Nielsen, Michael A and Chuang, Isaac L},
  year={2010},
  publisher={Cambridge University Press, Cambridge}
}

@article{Schuster2007,
  title = {Resolving photon number states in a superconducting circuit},
  volume = {445},
  ISSN = {1476-4687},
  url = {http://dx.doi.org/10.1038/nature05461},
  DOI = {10.1038/nature05461},
  number = {7127},
  journal = {Nature (London)},
  publisher = {Springer Science and Business Media LLC},
  author = {Schuster,  D. I. and Houck,  A. A. and Schreier,  J. A. and Wallraff,  A. and Gambetta,  J. M. and Blais,  A. and Frunzio,  L. and Majer,  J. and Johnson,  B. and Devoret,  M. H. and Girvin,  S. M. and Schoelkopf,  R. J.},
  year = {2007},
  month = feb,
  pages = {515}
}

@inproceedings{Grover1996,
    author = {Grover, Lov K.},
    title = {A fast quantum mechanical algorithm for database search},
    year = {1996},
    isbn = {0897917855},
    publisher = {Association for Computing Machinery},
    address = {New York},
    url = {https://doi.org/10.1145/237814.237866},
    doi = {10.1145/237814.237866},
    booktitle = {Proceedings of the Twenty-Eighth Annual ACM Symposium on Theory of Computing},
    pages = {212–219},
    numpages = {8},
    location = {Philadelphia, Pennsylvania, USA}
    }

@article{Brassard2002,
  title = {Quantum amplitude amplification and estimation},
  DOI = {10.1090/conm/305},
  journal = {Contemp. Math.},
  author = {Brassard,  Gilles and Høyer,  Peter and Mosca,  Michele and Tapp,  Alain},
  year = {2002},
  volume = {305},
  pages = {53}
}

@article{BlaisRMP2021,
  title = {Circuit quantum electrodynamics},
  author = {Blais, Alexandre and Grimsmo, Arne L. and Girvin, Steven M. and Wallraff, Andreas},
  journal = {Rev. Mod. Phys.},
  volume = {93},
  pages = {025005},
  year = {2021},
  doi = {10.1103/RevModPhys.93.025005}
}

@article{JaynesCummings1963,
  title = {Comparison of quantum and semiclassical radiation theories with application to the beam maser},
  author = {Jaynes, E. T. and Cummings, F. W.},
  journal = {Proc. IEEE},
  volume = {51},
  number = {1},
  pages = {89--109},
  year = {1963},
  doi = {10.1109/PROC.1963.1664}
}

@article{PhysRevLett.78.2547,
  title = {Method for Direct Measurement of the {Wigner} Function in Cavity {QED} and Ion Traps},
  author = {Lutterbach, L. G. and Davidovich, L.},
  journal = {Phys. Rev. Lett.},
  volume = {78},
  issue = {13},
  pages = {2547},
  numpages = {0},
  year = {1997},
  month = {Mar},
  publisher = {American Physical Society},
  doi = {10.1103/PhysRevLett.78.2547},
  url = {https://link.aps.org/doi/10.1103/PhysRevLett.78.2547}
}

@article{Bertet2002,
  title = {Direct Measurement of the {Wigner} Function of a One-Photon {Fock} State in a Cavity},
  author = {Bertet, P. and Auffeves, A. and Maioli, P. and Osnaghi, S. and Meunier, T. and Brune, M. and Raimond, J. M. and Haroche, S.},
  journal = {Phys. Rev. Lett.},
  volume = {89},
  issue = {20},
  pages = {200402},
  numpages = {4},
  year = {2002},
  month = {Oct},
  publisher = {American Physical Society},
  doi = {10.1103/PhysRevLett.89.200402},
  url = {https://link.aps.org/doi/10.1103/PhysRevLett.89.200402}
}

@article{Travaglione2001,
  title = {Generation of eigenstates using the phase-estimation algorithm},
  author = {Travaglione, B. C. and Milburn, G. J.},
  journal = {Phys. Rev. A},
  volume = {63},
  issue = {3},
  pages = {032301},
  numpages = {5},
  year = {2001},
  month = {Feb},
  publisher = {American Physical Society},
  doi = {10.1103/PhysRevA.63.032301},
  url = {https://link.aps.org/doi/10.1103/PhysRevA.63.032301}
}

@article{Delglise2008,
  title = {Reconstruction of non-classical cavity field states with snapshots of their decoherence},
  volume = {455},
  ISSN = {1476-4687},
  url = {http://dx.doi.org/10.1038/nature07288},
  DOI = {10.1038/nature07288},
  number = {7212},
  journal = {Nature (London)},
  publisher = {Springer Science and Business Media LLC},
  author = {Deléglise,  Samuel and Dotsenko,  Igor and Sayrin,  Clément and Bernu,  Julien and Brune,  Michel and Raimond,  Jean-Michel and Haroche,  Serge},
  year = {2008},
  month = sep,
  pages = {510}
}

@book{Quantum_HSJ,
    author = {Haroche, Serge and Raimond, Jean-Michel},
    title = {Exploring the Quantum: Atoms, Cavities, and Photons},
    publisher = {Oxford University Press},
    year = {2006},
    month = {08},
    isbn = {9780198509141},
    doi = {10.1093/acprof:oso/9780198509141.001.0001},
    url = {https://doi.org/10.1093/acprof:oso/9780198509141.001.0001},
}

@article{Rivera2023,
  title = {Creating large {Fock} states and massively squeezed states in optics using systems with nonlinear bound states in the continuum},
  volume = {120},
  ISSN = {1091-6490},
  url = {http://dx.doi.org/10.1073/pnas.2219208120},
  DOI = {10.1073/pnas.2219208120},
  number = {9},
  journal = {Proc. Natl. Acad. Sci. USA},
  publisher = {Proceedings of the National Academy of Sciences},
  pages = {e2219208120},
  author = {Rivera,  Nicholas and Sloan,  Jamison and Salamin,  Yannick and Joannopoulos,  John D. and Soljačić,  Marin},
  year = {2023},
  month = feb 
}

@article{PhysRevA.110.042421,
  title = {Generating {Fock}-state superpositions from coherent states by selective measurement},
  author = {Zhang, Chen-yi and Jing, Jun},
  journal = {Phys. Rev. A},
  volume = {110},
  issue = {4},
  pages = {042421},
  numpages = {12},
  year = {2024},
  month = {Oct},
  publisher = {American Physical Society},
  doi = {10.1103/PhysRevA.110.042421},
  url = {https://link.aps.org/doi/10.1103/PhysRevA.110.042421}
}

@article{Heeres2015,
  title = {Cavity State Manipulation Using Photon-Number Selective Phase Gates},
  author = {Heeres, Reinier W. and Vlastakis, Brian and Holland, Eric and Krastanov, Stefan and Albert, Victor V. and Frunzio, Luigi and Jiang, Liang and Schoelkopf, Robert J.},
  journal = {Phys. Rev. Lett.},
  volume = {115},
  issue = {13},
  pages = {137002},
  numpages = {5},
  year = {2015},
  month = {Sep},
  publisher = {American Physical Society},
  doi = {10.1103/PhysRevLett.115.137002},
  url = {https://link.aps.org/doi/10.1103/PhysRevLett.115.137002}
}

@article{Chen2024,
  title = {High-dimensional two-photon quantum controlled phase-flip gate},
  author = {Chen, Mingyuan and Tang, Jiang-Shan and Cai, Miao and Lu, Yanqing and Nori, Franco and Xia, Keyu},
  journal = {Phys. Rev. Res.},
  volume = {6},
  issue = {3},
  pages = {033004},
  numpages = {13},
  year = {2024},
  month = {Jul},
  publisher = {American Physical Society},
  doi = {10.1103/PhysRevResearch.6.033004},
  url = {https://link.aps.org/doi/10.1103/PhysRevResearch.6.033004}
}

@article{Glaser2023,
  title = {Controlled-Controlled-Phase Gates for Superconducting Qubits Mediated by a Shared Tunable Coupler},
  author = {Glaser, Niklas J. and Roy, Federico and Filipp, Stefan},
  journal = {Phys. Rev. Appl.},
  volume = {19},
  issue = {4},
  pages = {044001},
  numpages = {13},
  year = {2023},
  month = {Apr},
  publisher = {American Physical Society},
  doi = {10.1103/PhysRevApplied.19.044001},
  url = {https://link.aps.org/doi/10.1103/PhysRevApplied.19.044001}
}

@article{Barenco1995,
  title = {Elementary gates for quantum computation},
  author = {Barenco, Adriano and Bennett, Charles H. and Cleve, Richard and DiVincenzo, David P. and Margolus, Norman and Shor, Peter and Sleator, Tycho and Smolin, John A. and Weinfurter, Harald},
  journal = {Phys. Rev. A},
  volume = {52},
  issue = {5},
  pages = {3457},
  numpages = {0},
  year = {1995},
  month = {Nov},
  publisher = {American Physical Society},
  doi = {10.1103/PhysRevA.52.3457},
  url = {https://link.aps.org/doi/10.1103/PhysRevA.52.3457}
}

@article{Wallraff2004,
  title = {Strong coupling of a single photon to a superconducting qubit using circuit quantum electrodynamics},
  volume = {431},
  ISSN = {1476-4687},
  url = {http://dx.doi.org/10.1038/nature02851},
  DOI = {10.1038/nature02851},
  number = {7005},
  journal = {Nature (London)},
  publisher = {Springer Science and Business Media LLC},
  author = {Wallraff,  A. and Schuster,  D. I. and Blais,  A. and Frunzio,  L. and Huang,  R.- S. and Majer,  J. and Kumar,  S. and Girvin,  S. M. and Schoelkopf,  R. J.},
  year = {2004},
  month = sep,
  pages = {162}
}

@article{Kjaergaard2020,
  title = {Superconducting Qubits: Current State of Play},
  volume = {11},
  ISSN = {1947-5462},
  url = {http://dx.doi.org/10.1146/annurev-conmatphys-031119-050605},
  DOI = {10.1146/annurev-conmatphys-031119-050605},
  number = {1},
  journal = { Annu. Rev. Condens. Matter Phys. },
  publisher = {Annual Reviews},
  author = {Kjaergaard,  Morten and Schwartz,  Mollie E. and Braum\"{u}ller,  Jochen and Krantz,  Philip and Wang,  Joel I.-J. and Gustavsson,  Simon and Oliver,  William D.},
  year = {2020},
  month = mar,
  pages = {369}
}

@article{Krantz2019,
  author = {Krantz, P. and Kjaergaard, M. and Yan, F. and Orlando, T. P. and Gustavsson, S. and Oliver, W. D.},
    title = {A quantum engineer's guide to superconducting qubits},
    journal = {Appl. Phys. Rev.},
    volume = {6},
    number = {2},
    pages = {021318},
    year = {2019},
    month = {06},
    issn = {1931-9401},
    doi = {10.1063/1.5089550},
    url = {https://doi.org/10.1063/1.5089550}
}

@article{Molmer93,
author = {Klaus M{\o}lmer and Yvan Castin and Jean Dalibard},
journal = {J. Opt. Soc. Am. B},
number = {3},
pages = {524},
publisher = {Optica Publishing Group},
title = {Monte {Carlo} wave-function method in quantum optics},
volume = {10},
month = {Mar},
year = {1993},
url = {https://opg.optica.org/josab/abstract.cfm?URI=josab-10-3-524},
doi = {10.1364/JOSAB.10.000524}
}

@article{Johansson2012,
title = {{QuTiP}: An open-source Python framework for the dynamics of open quantum systems},
journal = {Comput. Phys. Commun.},
volume = {183},
number = {8},
pages = {1760},
year = {2012},
issn = {0010-4655},
doi = {https://doi.org/10.1016/j.cpc.2012.02.021},
url = {https://www.sciencedirect.com/science/article/pii/S0010465512000835},
author = {J.R. Johansson and P.D. Nation and Franco Nori}
}

@book{breuer2002theory,
  title={The Theory of Open Quantum Systems},
  author={Breuer, Heinz-Peter and Petruccione, Francesco},
  year={2002},
  publisher={Oxford University Press, Oxford}
}

@misc{li2026,
      title={Scalable Generation of Macroscopic {Fock} States Exceeding 10,000 Photons}, 
      author={Ming Li and Weizhou Cai and Ziyue Hua and Yifang Xu and Yilong Zhou and Zi-Jie Chen and Xu-Bo Zou and Guang-Can Guo and Luyan Sun and Chang-Ling Zou},
      year={2026},
      eprint={2601.05118},
      archivePrefix={arXiv},
      primaryClass={quant-ph},
      url={https://arxiv.org/abs/2601.05118}, 
}

@misc{ZhangJing2026,
  title = {Generating {Fock} state exceeding 10000 excitations with near unit fidelity by adaptive generalized-parity measurement},
  author = {Zhang, Chen-yi and Jing, Jun},
  year = {2026},
  eprint = {2606.01341},
  archivePrefix = {arXiv},
  primaryClass = {quant-ph},
  url = {https://arxiv.org/abs/2606.01341}
}

@misc{Austin2026,
  title = {{Fock}-state preparation based on amplitude amplification in cavity {QED}},
  author = {Austin, Sharoon and Wei, Zhi-Yuan and Srinivasan, Kartik and Gorshkov, Alexey V.},
  year = {2026},
  eprint = {2607.14239},
  archivePrefix = {arXiv},
  primaryClass = {quant-ph},
  url = {https://arxiv.org/abs/2607.14239}
}

@article{Teja2023,
  title = {Distillation of optical Fock states using atom-cavity systems},
  author = {Teja, G.P. and Chanchal},
  journal = {Phys. Rev. Appl.},
  volume = {20},
  issue = {4},
  pages = {044049},
  numpages = {8},
  year = {2023},
  month = {Oct},
  publisher = {American Physical Society},
  doi = {10.1103/PhysRevApplied.20.044049},
  url = {https://link.aps.org/doi/10.1103/PhysRevApplied.20.044049}
}

@misc{Jin2026,
  title = {Deterministic Generation of Arbitrary {Fock} States via Resonant Subspace Engineering},
  author = {Jin, Shan and Li, Ming and Cai, Weizhou and Chen, Zi-Jie and Xu, Yifang and Zhou, Yilong and Huang, Hongwei and Zhu, Yunlai and Hua, Ziyue and Guo, Guang-Can and Sun, Luyan and Wang, Xiaoting and Zou, Chang-Ling},
  year = {2026},
  eprint = {2602.12156},
  archivePrefix = {arXiv},
  primaryClass = {quant-ph},
  url = {https://arxiv.org/abs/2602.12156}
}

@article{Xu2026,
  title = {Principles of optics in {Fock} space for the scalable manipulation of large quantum states},
  author = {Xu, Yifang and Zhou, Yilong and Hua, Ziyue and Sun, Lida and Zhou, Jie and Wang, Weiting and Cai, Weizhou and Huang, Hongwei and Xiao, Lintao and Xue, Guangming and Yu, Haifeng and Li, Ming and Zou, Chang-Ling and Sun, Luyan},
  journal = {Nat. Phys.},
  year = {2026},
  publisher={Nature Publishing Group UK London},
  doi = {10.1038/s41567-026-03370-9},
  url = {https://doi.org/10.1038/s41567-026-03370-9}
}

@article{HollandBurnett1993,
  title   = {Interferometric Detection of Optical Phase Shifts at the {Heisenberg} Limit},
  author  = {Holland, M. J. and Burnett, K.},
  journal = {Phys. Rev. Lett.},
  volume  = {71},
  number  = {9},
  pages   = {1355},
  year    = {1993},
  doi     = {10.1103/PhysRevLett.71.1355}
}

@article{Gao2019,
  title={Entanglement of bosonic modes through an engineered exchange interaction},
  author={Gao, Yvonne Y and Lester, Brian J and Chou, Kevin S and Frunzio, Luigi and Devoret, Michel H and Jiang, Liang and Girvin, SM and Schoelkopf, Robert J},
  journal={Nature (London)},
  volume={566},
  number={7745},
  pages={509--512},
  year={2019},
  publisher={Nature Publishing Group UK London}
}

@article{GottesmanKitaevPreskill2001,
  title   = {Encoding a Qubit in an Oscillator},
  author  = {Gottesman, Daniel and Kitaev, Alexei and Preskill, John},
  journal = {Phys. Rev. A},
  volume  = {64},
  pages   = {012310},
  year    = {2001},
  doi     = {10.1103/PhysRevA.64.012310}
}

@article{Ofek2016Nature,
  title   = {Extending the lifetime of a quantum bit with error correction in superconducting circuits},
  author  = {Ofek, Nissim and Petrenko, Andrei and Heeres, Reinier W. and Reinhold, Philip and Leghtas, Zaki and Vlastakis, Brian and Liu, Yehan and Frunzio, Luigi and Girvin, S. M. and Jiang, Liang and Mirrahimi, Mazyar and Devoret, M. H. and Schoelkopf, R. J.},
  journal = {Nature (London)},
  volume  = {536},
  pages   = {441},
  year    = {2016},
  doi     = {10.1038/nature18949}
}

@article{CampagneIbarcq2020,
  title   = {Quantum error correction of a qubit encoded in grid states of an oscillator},
  author  = {Campagne-Ibarcq, P. and Eickbusch, A. and Touzard, S. and Zalys-Geller, E. and Frattini, N. E. and Sivak, V. V. and Reinhold, P. and Puri, S. and Shankar, S. and Schoelkopf, R. J. and Frunzio, L. and Mirrahimi, M. and Devoret, M. H.},
  journal = {Nature (London)},
  volume  = {584},
  pages   = {368},
  year    = {2020},
  doi     = {10.1038/s41586-020-2603-3}
}

@article{Huver2008,
  title = {Entangled {Fock} states for robust quantum optical metrology, imaging, and sensing},
  author = {Huver, Sean D. and Wildfeuer, Christoph F. and Dowling, Jonathan P.},
  journal = {Phys. Rev. A},
  volume = {78},
  issue = {6},
  pages = {063828},
  numpages = {5},
  year = {2008},
  month = {Dec},
  publisher = {American Physical Society},
  doi = {10.1103/PhysRevA.78.063828},
  url = {https://link.aps.org/doi/10.1103/PhysRevA.78.063828}
}

@article{Krastanov2015,
  title = {Universal control of an oscillator with dispersive coupling to a qubit},
  author = {Krastanov, Stefan and Albert, Victor V. and Shen, Chao and Zou, Chang-Ling and Heeres, Reinier W. and Vlastakis, Brian and Schoelkopf, Robert J. and Jiang, Liang},
  journal = {Phys. Rev. A},
  volume = {92},
  issue = {4},
  pages = {040303},
  numpages = {5},
  year = {2015},
  month = {Oct},
  publisher = {American Physical Society},
  doi = {10.1103/PhysRevA.92.040303},
  url = {https://link.aps.org/doi/10.1103/PhysRevA.92.040303}
}

@article{damas2025,
  title = {Engineered {Kerr} nonlinearities for precise quantum control of {Fock} states},
  author = {Damas, Gabriella G. and Diniz, Ciro Micheletti and de Almeida, Norton G. and Villas-B\^oas, Celso J. and de Moraes Neto, G. D.},
  journal = {Phys. Rev. Appl.},
  volume = {25},
  issue = {3},
  pages = {034097},
  numpages = {17},
  year = {2026},
  month = {Mar},
  publisher = {American Physical Society},
 doi = {10.1103/2q95-sfjs},
url = {https://link.aps.org/doi/10.1103/2q95-sfjs}
}

@article{Torosov2009,
  title = {Design of quantum {Fourier} transforms and quantum algorithms by using circulant {Hamiltonians}},
  author = {Torosov, Boyan T. and Vitanov, Nikolay V.},
  journal = {Phys. Rev. A},
  volume = {80},
  issue = {2},
  pages = {022329},
  numpages = {5},
  year = {2009},
  month = {Aug},
  publisher = {American Physical Society},
  doi = {10.1103/PhysRevA.80.022329},
  url = {https://link.aps.org/doi/10.1103/PhysRevA.80.022329}
}

@article{cleve1998quantum,
  title={Quantum algorithms revisited},
  author={Cleve, Richard and Ekert, Artur and Macchiavello, Chiara and Mosca, Michele},
  journal={Proc. R. Soc. London. Ser. A: Math., Phys. Eng. Sci.},
  volume={454},
  number={1969},
  pages={339},
  year={1998},
  publisher={The Royal Society},
  doi={10.1098/rspa.1998.0164}
}

@book{AbramowitzStegun,
  editor    = {Abramowitz, Milton and Stegun, Irene A.},
  title     = {Handbook of Mathematical Functions with Formulas, Graphs, and Mathematical Tables},
  publisher = {Dover},
  address   = {New York},
  year      = {1964},
}

@article{Reagor2016,
  title = {Quantum memory with millisecond coherence in circuit QED},
  author = {Reagor, Matthew and Pfaff, Wolfgang and Axline, Christopher and Heeres, Reinier W. and Ofek, Nissim and Sliwa, Katrina and Holland, Eric and Wang, Chen and Blumoff, Jacob and Chou, Kevin and Hatridge, Michael J. and Frunzio, Luigi and Devoret, Michel H. and Jiang, Liang and Schoelkopf, Robert J.},
  journal = {Phys. Rev. B},
  volume = {94},
  issue = {1},
  pages = {014506},
  numpages = {8},
  year = {2016},
  month = {Jul},
  publisher = {American Physical Society},
  doi = {10.1103/PhysRevB.94.014506},
  url = {https://link.aps.org/doi/10.1103/PhysRevB.94.014506}
}

@article{Milul2023,
  title = {Superconducting Cavity Qubit with Tens of Milliseconds Single-Photon Coherence Time},
  author = {Milul, Ofir and Guttel, Barkay and Goldblatt, Uri and Hazanov, Sergey and Joshi, Lalit M. and Chausovsky, Daniel and Kahn, Nitzan and \ifmmode \mbox{\c{C}}\else \c{C}\fi{}ifty\"urek, Engin and Lafont, Fabien and Rosenblum, Serge},
  journal = {PRX Quantum},
  volume = {4},
  issue = {3},
  pages = {030336},
  numpages = {16},
  year = {2023},
  month = {Sep},
  publisher = {American Physical Society},
  doi = {10.1103/PRXQuantum.4.030336},
  url = {https://link.aps.org/doi/10.1103/PRXQuantum.4.030336}
}

@article{Wagner2000,
doi = {10.1088/1464-4266/2/3/314},
url = {https://doi.org/10.1088/1464-4266/2/3/314},
year = {2000},
month = {jun},
publisher = {},
volume = {2},
number = {3},
pages = {306},
author = {Wagner Duarte José and Salomon S Mizrahi},
title = {Generation of circular states and {Fock} states in a
trapped ion},
journal = {J. Opt. B: Quantum Semiclass. Opt.}
}

@article{Ragi2000,
  title = {Non-classical properties of even circular states},
  author = {Ragi, R. and Baseia, B. and Mizrahi, S. S.},
  journal = {J. Opt. B: Quantum Semiclass. Opt.},
  volume = {2},
  number = {3},
  pages = {299--305},
  year = {2000},
  doi = {10.1088/1464-4266/2/3/313},
  url = {https://doi.org/10.1088/1464-4266/2/3/313}
}

@article{Heras2024,
  title = {Photonic quantum metrology with variational quantum optical nonlinearities},
  author = {Mu\~noz de las Heras, A. and Tabares, C. and Schneider, J. T. and Tagliacozzo, L. and Porras, D. and Gonz\'alez-Tudela, A.},
  journal = {Phys. Rev. Res.},
  volume = {6},
  issue = {1},
  pages = {013299},
  numpages = {14},
  year = {2024},
  month = {Mar},
  publisher = {American Physical Society},
  doi = {10.1103/PhysRevResearch.6.013299},
  url = {https://link.aps.org/doi/10.1103/PhysRevResearch.6.013299}
}

@article{SanchezMunoz18,
author = {Sánchez Muñoz,  Carlos and Laussy,  Fabrice P. and del Valle,  Elena and Tejedor,  Carlos and González-Tudela,  Alejandro},
journal = {Optica},
number = {1},
pages = {14},
publisher = {Optica Publishing Group},
title = {Filtering multiphoton emission from state-of-the-art cavity quantum electrodynamics},
volume = {5},
month = {Jan},
year = {2018},
url = {https://opg.optica.org/optica/abstract.cfm?URI=optica-5-1-14},
doi = {10.1364/OPTICA.5.000014}
}

@article{Tudela2015,
  title = {Deterministic Generation of Arbitrary Photonic States Assisted by Dissipation},
  author = {Gonz\'alez-Tudela, A. and Paulisch, V. and Chang, D. E. and Kimble, H. J. and Cirac, J. I.},
  journal = {Phys. Rev. Lett.},
  volume = {115},
  issue = {16},
  pages = {163603},
  numpages = {6},
  year = {2015},
  month = {Oct},
  publisher = {American Physical Society},
  doi = {10.1103/PhysRevLett.115.163603},
  url = {https://link.aps.org/doi/10.1103/PhysRevLett.115.163603}
}

@article{Tudela2017,
  title = {Efficient Multiphoton Generation in Waveguide Quantum Electrodynamics},
  author = {Gonz\'alez-Tudela, A. and Paulisch, V. and Kimble, H. J. and Cirac, J. I.},
  journal = {Phys. Rev. Lett.},
  volume = {118},
  issue = {21},
  pages = {213601},
  numpages = {6},
  year = {2017},
  month = {May},
  publisher = {American Physical Society},
  doi = {10.1103/PhysRevLett.118.213601},
  url = {https://link.aps.org/doi/10.1103/PhysRevLett.118.213601}
}

@article{Allman2010,
  title = {rf-SQUID-Mediated Coherent Tunable Coupling between a Superconducting Phase Qubit and a Lumped-Element Resonator},
  author = {Allman, M. S. and Altomare, F. and Whittaker, J. D. and Cicak, K. and Li, D. and Sirois, A. and Strong, J. and Teufel, J. D. and Simmonds, R. W.},
  journal = {Phys. Rev. Lett.},
  volume = {104},
  issue = {17},
  pages = {177004},
  numpages = {4},
  year = {2010},
  month = {Apr},
  publisher = {American Physical Society},
  doi = {10.1103/PhysRevLett.104.177004},
  url = {https://link.aps.org/doi/10.1103/PhysRevLett.104.177004}
}

@article{wulschner2016,
  title={Tunable coupling of transmission-line microwave resonators mediated by an rf {SQUID}},
  author = {Wulschner,  Friedrich and Goetz,  Jan and Koessel,  Fabian R and Hoffmann,  Elisabeth and Baust,  Alexander and Eder,  Peter and Fischer,  Michael and Haeberlein,  Max and Schwarz,  Manuel J and Pernpeintner,  Matthias and Xie,  Edwar and Zhong,  Ling and Zollitsch,  Christoph W and Peropadre,  Borja and Garcia Ripoll,  Juan-Jose and Solano,  Enrique and Fedorov,  Kirill G and Menzel,  Edwin P and Deppe,  Frank and Marx,  Achim and Gross,  Rudolf},
  journal={EPJ Quantum Techno.},
  volume={3},
  number={1},
  pages={10},
  year={2016},
  publisher={Springer},
   url = {http://dx.doi.org/10.1140/epjqt/s40507-016-0048-2},
  DOI = {10.1140/epjqt/s40507-016-0048-2}
}

@article{Boto2000QuantumLithography,
  title = {Quantum Interferometric Optical Lithography: Exploiting Entanglement to Beat the Diffraction Limit},
  author = {Boto, Agedi N. and Kok, Pieter and Abrams, Daniel S. and Braunstein, Samuel L. and Williams, Colin P. and Dowling, Jonathan P.},
  journal = {Phys. Rev. Lett.},
  volume = {85},
  pages = {2733},
  year = {2000},
  doi = {10.1103/PhysRevLett.85.2733}
}

@article{Dowling2008HighNOON,
  title = {Quantum optical metrology---the lowdown on high-{N00N} states},
  author = {Dowling, Jonathan P.},
  journal = {Contemp. Phys.},
  volume = {49},
  number = {2},
  pages = {125},
  year = {2008},
  doi = {10.1080/00107510802091298}
}

@article{Lu2023HighFidelityBeamsplitter,
  title = {High-fidelity parametric beamsplitting with a parity-protected converter},
  author = {Lu, Yao and Maiti, Aniket and Garmon, John W. O. and Ganjam, Suhas and Zhang, Yaxing and Claes, Jahan and Frunzio, Luigi and Girvin, S. M. and Schoelkopf, Robert J.},
  journal = {Nat. Commun.},
  volume = {14},
  pages = {5767},
  year = {2023},
  doi = {10.1038/s41467-023-41104-0}
}

@article{Su2014NOONCircuitQED,
  title = {Fast and simple scheme for generating {NOON} states of photons in circuit {QED}},
  author = {Su, Qi-Ping and Yang, Chui-Ping and Zheng, Shi-Biao},
  journal = {Sci. Rep.},
  volume = {4},
  pages = {3898},
  year = {2014},
  doi = {10.1038/srep03898}
}

@article{VillasBoas2005,
  title={One-step generation of high-quality squeezed and EPR states in cavity QED},
  author={Villas-Boas, CJ and Moussa, MHY},
  journal={The European Physical Journal D-Atomic, Molecular, Optical and Plasma Physics},
  volume={32},
  number={1},
  pages={147--151},
  year={2005},
  publisher={Springer},
  doi = {10.1140/epjd/e2004-00178-y}
}

@article{Prado2006,
  title = {Bilinear and quadratic Hamiltonians in two-mode cavity quantum electrodynamics},
  author = {Prado, F. O. and de Almeida, N. G. and Moussa, M. H. Y. and Villas-B\^oas, C. J.},
  journal = {Phys. Rev. A},
  volume = {73},
  issue = {4},
  pages = {043803},
  numpages = {5},
  year = {2006},
  month = {Apr},
  publisher = {American Physical Society},
  doi = {10.1103/PhysRevA.73.043803},
  url = {https://link.aps.org/doi/10.1103/PhysRevA.73.043803}
}

\end{document}